\documentclass[a4paper,11pt]{article}
\pdfoutput=1 % if your are submitting a pdflatex (i.e. if you have
\usepackage{jcappub} % for details on the use of the package, please
\usepackage[T1]{fontenc} % if needed
\usepackage{caption}
\usepackage{subcaption}
\usepackage{amsmath}

\newcommand{\be}{\begin{eqnarray}}
\newcommand{\ee}{\end{eqnarray}}

\title{Testing regular black holes\\with X-ray and GW data}

\author[a]{Shafqat~Riaz,}
\author[a]{Swarnim~Shashank,}
\author[a]{Rittick~Roy,}
\author[a,b,c]{Askar~B.~Abdikamalov,}
\author[d]{Dimitry~Ayzenberg,}
\author[a,1]{Cosimo~Bambi\note{Corresponding author.},}
\author[a]{Zuobin~Zhang,}
\author[e]{Menglei~Zhou}

\affiliation[a]{Center for Field Theory and Particle Physics and Department of Physics, Fudan University,\\
2005 Songhu Road, 200438 Shanghai, China}
\affiliation[b]{Ulugh Beg Astronomical Institute, Astronomy Str. 33, Tashkent 100052, Uzbekistan}
\affiliation[c]{Institute of Fundamental and Applied Research, National Research University TIIAME,\\
Kori Niyoziy 39, Tashkent 100000, Uzbekistan}
\affiliation[d]{Theoretical Astrophysics, Eberhard-Karls Universität Tübingen,\\
Auf der Morgenstelle 10, D-72076 Tübingen, Germany}
\affiliation[e]{Institut für Astronomie und Astrophysik, Eberhard-Karls Universität Tübingen,\\
Sand 1, D-72076 Tübingen, Germany}

\emailAdd{bambi@fudan.edu.cn}

\abstract{The presence of spacetime singularities in physically relevant solutions of the Einstein Equations is normally interpreted as a symptom of the breakdown of classical general relativity at very high densities/curvatures. However, despite significant efforts in the past decades, we do not have yet any robust theoretical framework to solve the problem of spacetime singularities. In this context, the past few years have seen an increasing interest in the study of phenomenological scenarios to describe singularity-free black holes, gravitational collapses, and cosmological models. In the present work, we consider the recent proposal by Mazza, Franzin \& Liberati for a rotating regular black hole and we measure their regularization parameter $l$ from the available X-ray and gravitational wave black hole data. For $l = 0$, we recover the singular Kerr solution of general relativity, while for $l \neq 0$ we can have a regular black hole or a regular wormhole. Our analysis shows that the available data are consistent with a vanishing regularization parameter $l$ and we can constrain its value. From a \textsl{NuSTAR} spectrum of the Galactic black hole in EXO~1846--031, we find $l/M < 0.49$ (90\% CL). From the gravitational wave event GW190707A, we find $l/M < 0.72$ (90\% CL).}

\begin{document}
\maketitle
\flushbottom

%%%%%%%%%%%%%%%%%%%%%%%%%%%%%%%%%%%%%%%%%%%%%%%%%%%%%

\section{Introduction}\label{sec:intro}

The general theory of relativity was proposed by Albert Einstein at the end of 1915~\cite{Einstein:1916vd}. The theory has been extremely successful: after more than 100~years and without any modification, it is still our standard framework for the description of the gravitational field and of the chrono-geometrical structure of the spacetime. The theory has passed a large number of observational tests, especially in the weak field regime, and all the available observational data are so far consistent with the theoretical predictions~\cite{Will:2014kxa}. However, there are a few theoretical issues that strongly suggest the existence of new physics and motivate current efforts to look for models beyond Einstein's gravity.

One of the most important problems of the general theory of relativity is the presence of spacetime singularities in some physically relevant solutions of the Einstein Equations\footnote{We note that there are a few different types of spacetime singularities. While there are specific relations among different types of spacetime singularities, here we mainly refer to geodesic singularities.}. At a singularity, predictability is lost and standard physics breaks down. While we do not have yet any robust theory of quantum gravity, it is commonly thought that the problem of spacetime singularities in the solutions of the Einstein Equations should be solved by quantizing gravity.

In the absence of any robust theory of quantum gravity, one can study the problem of spacetime singularities in quantum-gravity inspired models and explore the corresponding phenomenology. With this spirit, in the literature we can now find a number of proposals about regular black holes (see, e.g., Refs.~\cite{bardeen68,Dymnikova:1992ux,Ayon-Beato:1998hmi,Dymnikova:2003vt,Hayward:2005gi,Nicolini:2005vd,Ansoldi:2006vg,Spallucci:2008ez,Bambi:2013ufa,Frolov:2016pav,Bambi:2016wdn,Modesto:2021wdd}), singularity-free gravitational collapse models (see, e.g., Refs.~\cite{Frolov:1981mz,Bambi:2013caa,Bambi:2013gva,Bambi:2016uda}), and singularity-free cosmological models (see, e.g., Refs.~\cite{Bojowald:2001xe,Ashtekar:2006wn,Ashtekar:2006rx,Cai:2011tc,Qiu:2011cy,Alexander:2014eva}). While the first works along this line of research are now 40-50~years old, the past few years have really seen an increasing interest in this approach, with a remarkable number of new ideas and publications. All these models do not have a fundamental theory behind. They rather propose an effective metric in which the singularity problem is solved by violating some condition that holds in Einstein gravity in vacuum or in the presence of ordinary matter. It is remarkable that there are not so many qualitatively different scenarios if we still impose a few number of reasonable assumptions, and eventually we can classify all these models in a small number of groups~\cite{Carballo-Rubio:2019nel,Carballo-Rubio:2019fnb}.

In this work, we consider the metric recently proposed by Mazza, Franzin \& Liberati in Ref.~\cite{Mazza:2021rgq}, which is a rotating generalization of the Simpson-Visser metric~\cite{Simpson:2018tsi,Simpson:2019cer,Lobo:2020ffi}, and we infer observational constraints on its regularization parameter $l$. While there are potentially a few different methods to test the spacetime geometry around astrophysical black holes~\cite{Bambi:2015kza,Yagi:2016jml,Bambi:2017khi}, with the data available today the most stringent and robust constraints can be normally inferred from the analysis of the reflection features in the X-ray spectra of accreting black holes and from the analysis of the gravitational wave signal emitted by binary black holes at the coalescence~\cite{Tripathi:2020yts,Bambi:2021chr}. We will thus constrain the regularization parameter $l$ with these two methods. For the X-ray constraint, we will analyze a \textsl{NuSTAR} observation of the Galactic black hole EXO~1846--031 during its outburst in 2019. The source was particularly bright at that time and its spectrum had very strong reflection features, so it is an observation particularly suitable for testing the spacetime metric around black holes. For the gravitational wave constraint, we will fit the publicly available posterior samples released by the LIGO-Virgo Collaboration of the most suitable binary black hole events in GWTC-1 and GWTC-2. We note that the gravitational wave constraint is conservative because inferred from the modified motion of the binary system and assuming that the energy loss rate is the same as in general relativity. This is the only way to constrain the parameter $l$ because we only know the spacetime metric and we do not know the field equations of the underlying theory. In the case the energy loss rate introduced larger modifications in the gravitational wave signal, the actual constrain on $l$ would be stronger.

The manuscript is organized as follows. In Section~\ref{metric_overview}, we briefly review the metric proposed by Mazza, Franzin \& Liberati in Ref.~\cite{Mazza:2021rgq}. In Section~\ref{section-disk}, we construct two models to test black holes with X-ray data: the first model can be used to fit the thermal spectrum of the disk and the second model is to analyze the reflection component of the disk. In Section~\ref{sec:x}, we analyze a \textsl{NuSTAR} spectrum of the Galactic black hole in EXO~1846--031 and we constrain the regularization parameter $l$ of the Mazza-Franzin-Liberati metric from the analysis of the reflection features (the thermal component is too weak in this specific observation, so we cannot use the thermal model developed in Section~\ref{section-disk}). In Section~\ref{sec:gw}, we constrain the regularization parameter $l$ from the public measurements of the gravitational wave events released by the LIGO-Virgo Collaboration. Summary and conclusions are reported in Section~\ref{sec:dc}. In this paper, we always use natural units in which $G_{\rm N} = c = 1$ and a metric with signature $(-+++)$.

%%%%%%%%%%%%%%%%%%%%%%%%%%%%%%%%%%%%%%%%%%%%%%%%%%%%%

\section{Quick overview of the metric}\label{metric_overview}

This section provides a brief overview of the metrics used in our study. The spacetime geometry surrounding a rotating black hole ``mimicker'' was proposed in Ref.~\cite{Mazza:2021rgq}, as a rotating generalization of the Simpson-Visser metric~\cite{Simpson:2018tsi, Simpson:2019cer, Lobo:2020ffi}. The metric is stationary, axisymmetric, and asymptotically flat, and its line element in the Boyer-Lindquist coordinates ($t, r, \theta, \phi$) is written as~\cite{Mazza:2021rgq}
\begin{equation}
\label{line_element}
 {\rm d}s^2 = -\left(1 - \frac{2 M \sqrt{r^2 + l^2}}{\Sigma}\right){\rm d}t^2 + \frac{\Sigma}{\Delta}{\rm d}r^2 + \Sigma {\rm d}\theta^2 - \frac{4 M a \sin^2\theta \sqrt{r^2 + l^2}}{\Sigma}{\rm d}t{\rm d}\phi + \frac{A \sin^2\theta}{\Sigma}{\rm d}\phi^2 ,
\end{equation}
where
\begin{equation*}
\begin{gathered}
\Sigma = r^2 + l^2+ a^2 \cos^2\theta \, , \qquad
\Delta = r^2 + l^2 + a^2  - 2 M \sqrt{r^2 + l^2}  \,,
\\
A = \left(r^2+a^2 + l^2 \right)-\Delta a^2 \sin^2\theta \, .  
\end{gathered}
\end{equation*}
Here $M$ is the black hole mimicker mass, $a = J/M$ (dimensionless spin $a_{*} = a/M$), $J$ is the spin angular momentum, and $l \ge 0$ is the singularity regularization parameter and has the dimensions of length (in units of the gravitational radius $r_{\rm g} = M$).

Employing such a metric to analyze X-ray binary spectra is particularly compelling, as different values of $l$ and $a$ can indicate the presence of different types of black hole mimicker spacetimes. When $a = 0$, the above metric reduces to the Simpson-Visser metric, and a non-vanishing parameter $l$ gives the modification of the Schwarzschild black hole, while vanishing $l$ encompasses it. Similarly, we obtain the Kerr solution from the above metric for the special case of $l = 0$, and any non-vanishing value of $l$ is regarded as a deviation from the Kerr metric. It should be noted that the metric is not related to the Kerr solution by a change of coordinates. The metric~\ref{line_element} remains unchanged under a transformation of $r \rightarrow -r$ (i.e., a reflection symmetry), and, as a consequence, the spacetime geometry may be seen as being composed of two identical portions joined at $r = 0$. As long as $l \neq0$, the metric is regular everywhere, the Kretschmann scalar, $R^{\mu \nu \lambda \sigma} R_{\mu \nu \lambda \sigma}$, is finite (see appendix A in Ref.~\cite{Mazza:2021rgq}), the singularity is excised, and $r = 0$ is a regular surface (an oblate spheroid of size $l$), which may be crossed by an observer. The metric, thus, represents a wormhole spacetime with a throat of extent $l$ located at $r = 0$. The values of $l$ and $a$ indicate whether the throat is timelike, spacelike, or null. Note that throughout this study, only $a > 0$ is considered.

\begin{figure}[tbp]
\centering
\includegraphics[width=0.60\textwidth,trim=0 0 0 0,clip]{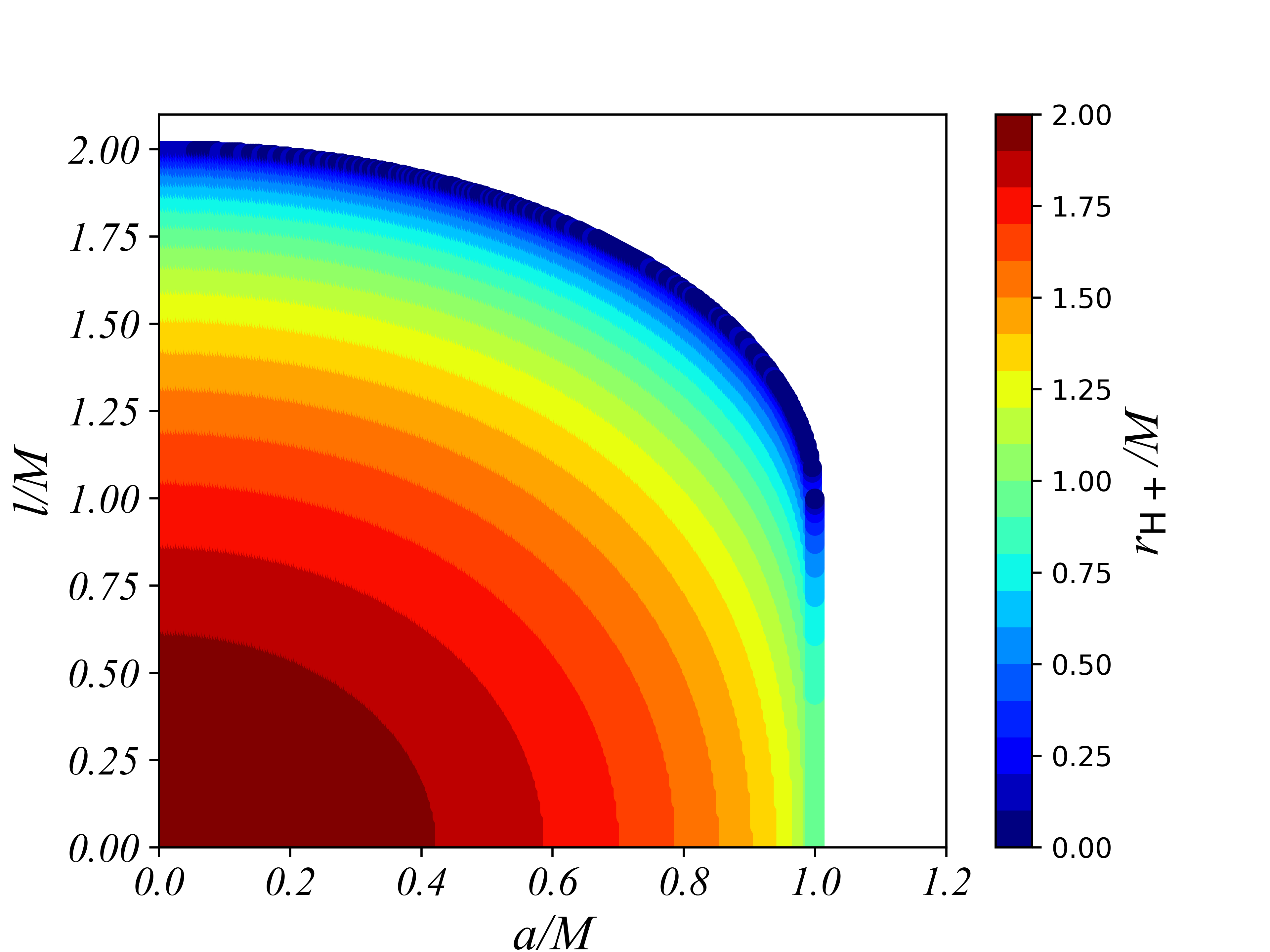}
\caption{Contour levels for the radial coordinate of the event horizon ($r_{\rm H+}$) as a function of $a$ and $l$ of the metric. There is no event horizon in the white region.}
\label{EHcontours}
\end{figure}

An interesting feature of the metric is that the existence of coordinate singularities depends on the values of $a$ and $l$. If these singularities exist, they can be found by setting $\Delta = 0$, which correspond to the horizons of the metric and are located at radial coordinates 
\begin{equation}
r_{\rm H \pm} = \sqrt{ \left( M \pm \sqrt{M^2 - a^2} \right)^2 - l^2 }, 
\end{equation}
where $r_{\rm H +}$ and $r_{\rm H -}$ denote, respectively, the outer (event) and the inner horizon. $r_{\rm H +}$ is real as long as $l \le M + \sqrt{M^2 - a^2}$ and $r_{\rm H -}$ is real as long as $l \le  M - \sqrt{M^2 - a^2}$. In the case of $l = 0$, we recover the horizons for the Kerr metric. $r_{\rm H+}$ is the event horizon, which varies from $2~M$ for $a = 0$ to $1~M$ for $a = M$, and now $a > M$ is no longer allowed since it results in a naked singularity. Figure~\ref{EHcontours} shows the contour levels of the event horizon for the metric~\ref{line_element} for different values of $a$ and $l$. The white region contains no coordinate singularities. The radius of the event horizon decreases with increasing both $a$ and $l$, which is quite intriguing.

\begin{figure}[tbp]
\centering 
\includegraphics[width=0.65\textwidth,trim=0 0 0 0,clip]{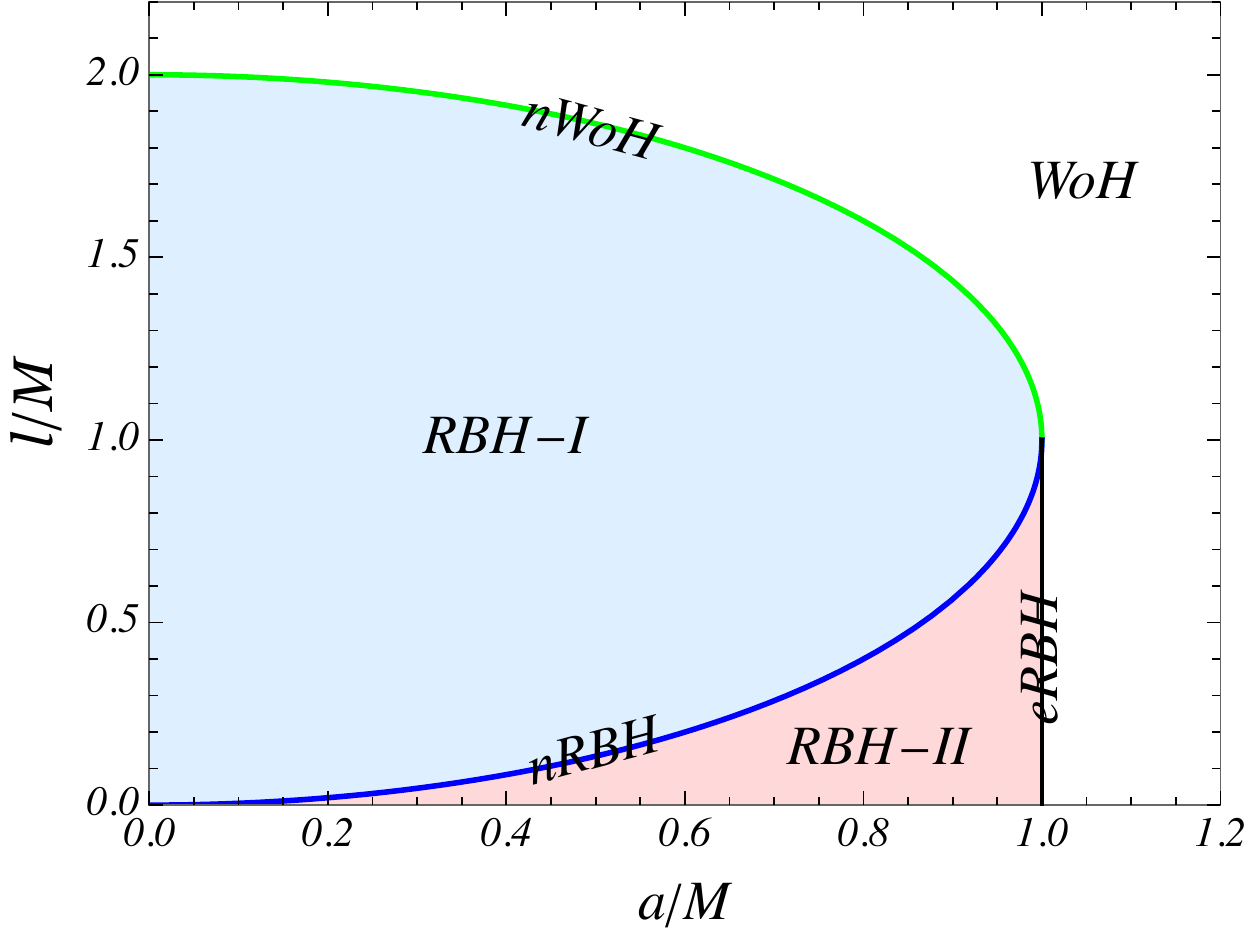}
\caption{Diagram indicating the different spacetime structures corresponding to different values of the parameters $a/M$ and $l/M$ of the metric. The acronyms are explained in the text. }
\label{phasediagram}
\end{figure}

Based on the values of $a$ and $l$, we may have the following six morphologies of the spacetime~\cite{Mazza:2021rgq} (see Figure~\ref{phasediagram}):
\begin{itemize}
    \item\textbf{WoH}: traversable wormhole, i.e., two-way wormhole with a timelike throat, which corresponds to either $a > M$ or $a < M$ and $l > M + \sqrt{M^2 - a^2}$; no singularity exists. This is indicated by the white region in Figure~\ref{phasediagram}. 
    
    \item \textbf{nWoH}: null wormhole with an extremal event horizon. It is possible to pass through the null throat one way. This is the case when $a < M$ and $l = M + \sqrt{M^2 - a^2}$. This is represented by the green curve in Figure~\ref{phasediagram}. 
    
    \item \textbf{RBH-I}: regular black hole that has one horizon per side; the singularity is excised by the throat of a spacelike wormhole. This is obtained when $a < M$ and $ M - \sqrt{M^2 - a^2} <l < M + \sqrt{M^2 - a^2}$. This is shown with the blue shaded region in Figure~\ref{phasediagram}.  
    
    \item \textbf{RBH-II}: regular black hole characterized by an inner and outer horizon for each of its sides, along with timelike throats associated with it. One can obtain this scenario by restricting $a < M$ and $l < M - \sqrt{M^2 - a^2}$. This is depicted by the red shaded area in Figure~\ref{phasediagram}. 
    
    \item \textbf{eRBH}: extremal regular black hole with an extremal horizon on each side. It describes the situation in which two horizons of a black hole coincide, corresponding to $a = M$ and $l < M$. This is shown with a dark vertical line in Figure~\ref{phasediagram}.  
    
    \item \textbf{nRBH}: regular black hole that possesses one horizon on each side and a null throat. It can be regarded as a limiting case of RBH-I and RBH-II, i.e., $a < M$ and $l = M - \sqrt{M^2 - a^2}$. This is represented by the blue curve in Figure~\ref{phasediagram}.  
\end{itemize}

\begin{figure}[tbp]
\centering 
\includegraphics[width=0.75\textwidth,trim=0 0 0 0,clip]{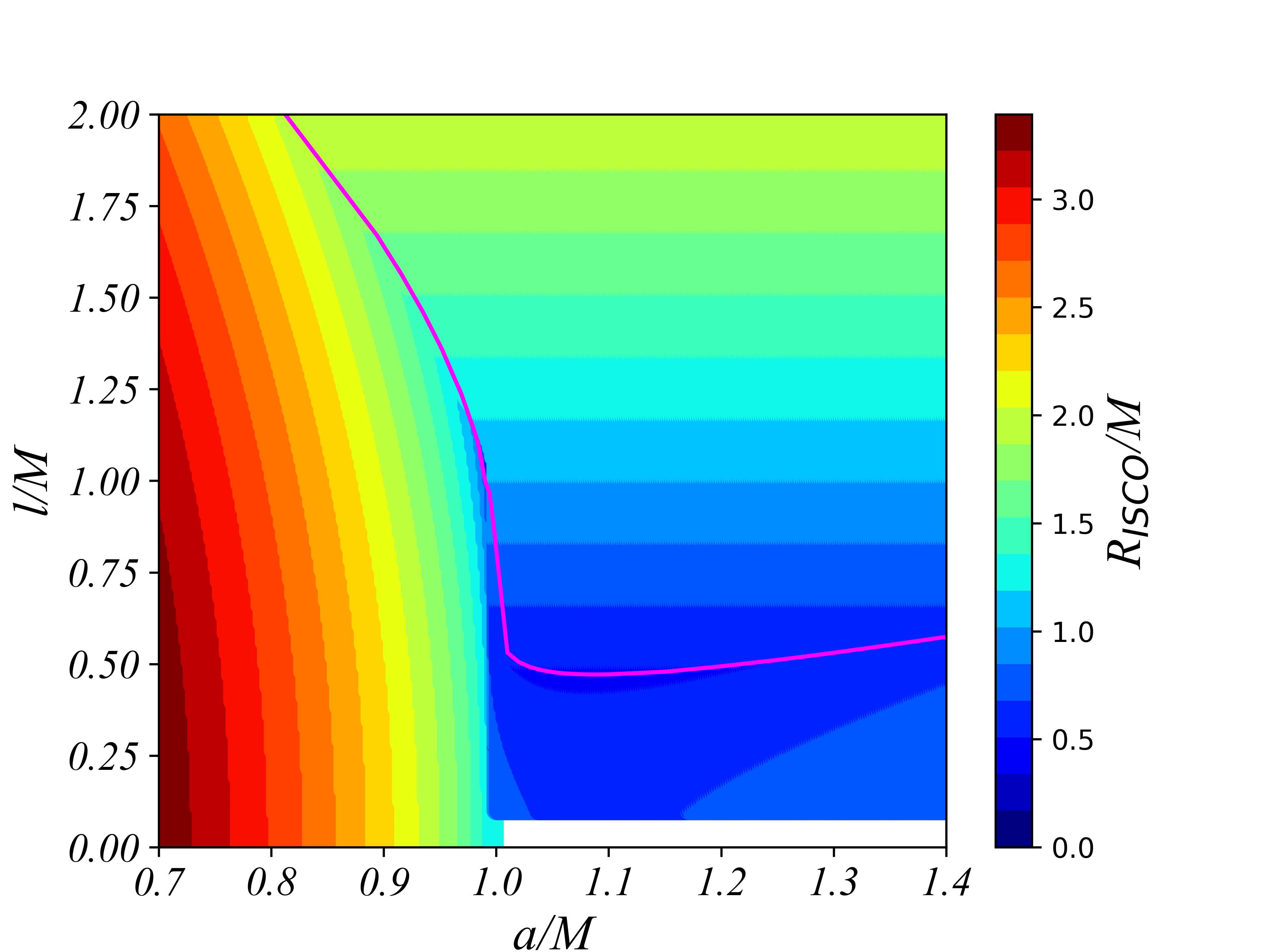}
\caption{Contour levels corresponding to the radial coordinate of ISCO for varying values of $a$ and $l$ of the metric. The magenta curve represents the scenario in which there is no event horizon, and the size of the wormhole throat corresponds to the radial coordinate of ISCO, i.e., $l = r_{\rm ISCO}$. For $a/M>1$, the parameter space with $l < 0.1~M$ is excluded in our analysis to avoid the region close to spacetimes with a naked singularity.}
\label{ISCOcontours}
\end{figure}

Another relevant and significant quantity is the radial coordinate of the innermost stable circular orbit (ISCO hereafter), $r_{\rm ISCO}$, in the equatorial plane. The derivation of $r_{\rm ISCO}$ for metric~\ref{line_element} is given in detail in Ref.~\cite{Mazza:2021rgq}, and thus is omitted here. In the case of $a \le M$ and $l = 0$, we recover $r_{\rm ISCO}$ of the Kerr metric, which is a monotonically decreasing function of the spin parameter. For example, when $a = -M$ (maximally counter-rotating), $r_{\rm ISCO} = 9~M$, when $a = 0$, it is $6~M$, and when $a = M$ (maximally rotating), it is $M$. Figure~\ref{ISCOcontours} shows the contour levels of $r_{\rm ISCO}$ for the metric~\ref{line_element}. Again, when $a > M$, $l = 0$ is excluded and the initial value of $l$ is set to 0.1~$~M$, as shown in the white region. This is in order to avoid complexities close to a naked singularity. The magenta curve represents the case when there is no event horizon, and the size of the wormhole coincides with $r_{\rm ISCO}$, meaning that $l = r_{\rm ISCO}$. Interestingly, below the magenta curve $r_{\rm ISCO}$ decreases with both spin and $l$ up to $a/M \sim 1.1$. It then exhibits an increasing trend with spin and a decreasing trend with $l$ for $a/M > 1.1$. On the other hand, above the magenta curve the contours are almost flat with the spin and increase with $l$. This is due to the fact that in this region we have wormholes with $r_{\rm ISCO}$ inside their throats, so we truncate the contours at a radial coordinate of $l + 0.01$, i.e., just outside the throat (Figure~\ref{ISCOcontours}). Note that for the ISCO contours, we consider $0.7< a/M < 1.4$, which is the spin parameter range adopted in the models; this will be discussed further in Section~\ref{model_nkbb}.

%%%%%%%%%%%%%%%%%%%%%%%%%%%%%%%%%%%%%%%%%%%%%%%%%%%%%

\section{Models for the electromagnetic spectrum of the accretion disk}\label{section-disk}

\subsection{Novikov-Thorne model}\label{ADM}

In this section, we briefly describe the accretion disk model employed in this work. In an astrophysical environment, accretion disk material around stellar-mass X-ray binaries comes from the companion star, whereas around galactic nuclei, it originates from interstellar gas~\cite{Bambi:2017iyh,Nampalliwar:2018tup}. The standard framework of accretion disks -- the Novikov-Thorne disk model~\cite{Page:1974he,Novikov:1973kta,Shakura:1972te} -- assumes that the accretion disk is geometrically thin, optically thick, and the gas particles follow nearly-geodesic circular orbits on the equatorial plane ($\theta = \pi/2$) of spacetime. In addition, the model requires that the spacetime is stationary, axisymmetric, and asymptotically flat. We assume that the energy is mainly radiated from the surface of the disk rather than heat transported in the radial direction. One can construct the time-averaged radial structure of the accretion disk by applying mass, energy, and angular momentum conservation. It is possible to express the energy flux radiated from the disk's surface as~\cite{Page:1974he,Novikov:1973kta}
\begin{equation}
\label{Eflux}
    \mathcal{F} = \frac{\dot{M}}{4 \pi M^2} F(r), 
\end{equation}
where $\dot{M} = dM/dt$ is the time-averaged mass accretion rate irrespective of the radial coordinate, $M$ is the mass of the black hole, and $F(r)$ is a dimensionless function which can be expressed as
\begin{equation}
\label{dimlF}
    F(r) = - \frac{\partial_{r}\Omega}{(E - \Omega L_{z} ^2)} \frac{M^2}{\sqrt{-G}} \int_{r_{\rm in}} ^ r  (E - \Omega L_{z}) (\partial_{\rho} L_{z}) d\rho, 
\end{equation}
where $G$ is the metric determinant on the near equatorial plane. $\Omega$, $L_{z}$, and $E$ are, respectively, the angular velocity of the geodesic circular orbits on the equatorial plane, the conserved axial component of the specific angular momentum, and the conserved specific energy. Ref.~\cite{Bambi:2017khi} provides the detailed derivation of $\Omega$, $L_z$, and $E$; here, however, we provide their expressions: 
\begin{equation}
\label{angular_velocity}
    \Omega = \frac{- \partial_{r} g_{t \phi} \pm  \sqrt{(\partial_{r} g_{t \phi})^2 - (\partial_{r} g_{tt}) (\partial_{r} g_{\phi \phi})}}{\partial_{r}g_{\phi \phi}}, 
\end{equation}
\begin{equation}
    L_{z} = \frac{g_{t \phi} + \Omega g_{\phi \phi}}{\sqrt{-g_{tt} - 2 \Omega g_{t\phi} - \Omega^2 g_{\phi \phi} }}, 
\end{equation}
\begin{equation}
    E = - \frac{g_{tt} + \Omega g_{t\phi}}{\sqrt{-g_{tt} - 2 \Omega g_{t\phi} - \Omega^2 g_{\phi \phi} }}, 
\end{equation}
where $+$ and $-$ in the expression of $\Omega$ represent, respectively, co-rotating and counter-rotating orbits, i.e., orbits with angular momentum parallel and anti-parallel to the black hole's spin. $g_{\mu \nu}$ are the components -- written in canonical form -- of a stationary, axisymmetric and asymptotically flat spacetime whose line element is given by
\begin{equation}
    ds^2 = g_{tt}dt^2 + 2g_{t\phi}dtd\phi + g_{rr}dr^2 + g_{\theta \theta}d{\theta}^2 + g_{\phi \phi}d{\phi ^2}.   
\end{equation}

In Eq.~\ref{dimlF}, $r_{\rm in}$ refers to inner edge of the accretion disk. The general consensus is that $r_{\rm in}$ stabilizes at $r_{\rm ISCO}$ in a thermally dominated state with a mass accretion rate ranging anywhere between 5\% and 30\% of the Eddington limit~\cite{Kulkarni:2011cy,Steiner:2010kd}. However, there is a possibility that the disk is truncated, that is, $r_{\rm in} > r_{\rm ISCO}$. The circular orbits within the ISCO radius are unstable; as a result, the gas material there plummets rapidly without emitting radiation and eventually crosses the event horizon. For the spacetime metric~\ref{line_element}, the inner edge of the accretion disk follows the contour levels shown in Figure~\ref{ISCOcontours}.

\begin{figure}[tbp]
\centering 
\includegraphics[width=0.75\textwidth,trim=0 0 0 0,clip]{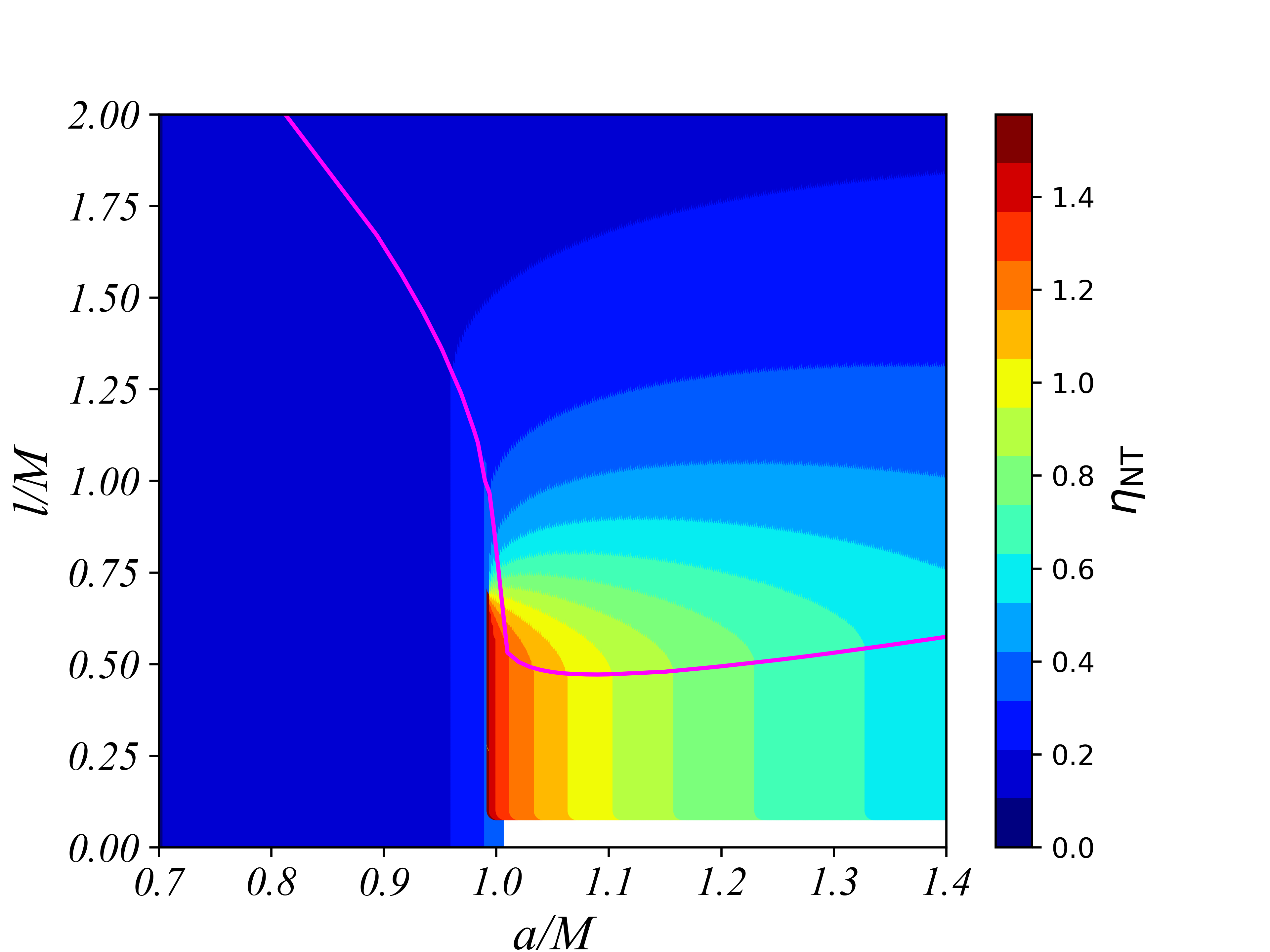}
\caption{Contour levels corresponding to the radiative efficiency of the Novikov-Thorne disk $(\eta_{\rm NT})$ for varying $a/M$ and $l/M$. The magenta curve represents the case when we have a wormhole, and its throat approaches the radial coordinate of ISCO orbit, i.e., $l = r_{\rm ISCO}$. The white region is excluded to avoid the complications close to a naked singularity.}
\label{eta_contours}
\end{figure}

Another crucial quantity associated with the Novikov-Thorne disk model is the radiative efficiency, defined as
\begin{equation}
    \eta_{\rm NT} = 1 - E_{\rm r_{\rm in}},
\end{equation}
where $E_{\rm r_{\rm in}}$ is the specific energy of massive particles (i.e., the energy normalized to the particle mass) at inner edge of the disk. The contour levels of the disk's radiative efficiency in the spacetime metric~\ref{line_element} are shown in Figure~\ref{eta_contours}. Interestingly, for $a/M \le 1$, the disk's radiative efficiency is almost constant with $l/M$ and only slightly varies with $a/M$. For $a/M > 1$, it exhibits a rapidly decreasing trend with $a/M$ under the magenta curve -- begins at a maximum around $a/M \sim 1$ -- which gradually becomes flatter as $l/M$ increases. Additionally, for $a/M > 1$, the radiative efficiency as a whole exhibits a decreasing trend as $l/M$ increases. Objects with $\eta_{\rm NT}$ greater than one are most likely unstable.

We note that the metric in Eq.~\ref{line_element} is not a vacuum solution of the Einstein Equations and therefore we have an effective non-vanishing energy-momentum tensor. Even if we interpret such a non-vanishing energy-momentum tensor as a new matter field, there are no modifications to the Novikov-Thorne model as long as the new matter field does not couple to the ordinary matter in the disk. The gas in the disk still follows the geodesics of the spacetime. The structure of the disk directly follows from the conservation of mass, energy, and angular momentum of the matter in the disk, which is not affected by the new matter field.

%%%%%%%%%%%%%%%%%%%%%%%%%%%%%%%%%%%%%%%%%%%%%%%%%%%%%

\subsection{Thermal spectrum model: {\tt nkbb}} \label{model_nkbb}

In this section, we construct a model for the thermal spectrum of the disk. More specifically, we implement the regular black hole metric in Eq.~\ref{line_element} in the XSPEC model \texttt{nkbb}~\cite{Zhou:2019fcg}, which is specifically designed to calculate the thermal spectrum of a thin accretion disk in non-Kerr spacetimes. The Novikov-Thorne disk model assumes that each point of the disk is in local thermal equilibrium and emits a spectrum like that of a blackbody. Since we are working with an axisymmetric spacetime, it is possible to derive an effective temperature of the disk as a function of radius $T_{\rm eff}(r)$ from the Stefan-Boltzmann law $F(r) = \sigma T^4_{\rm eff}$, where $\sigma$ is the Stefan-Boltzmann constant. The overall spectrum of the disk appears to be a multi-temperature blackbody spectrum~\cite{Li:2004aq,Bambi:2012tg,Bambi:2017khi}. Analyzing the thermal component of an accreting black hole system to determine its properties is commonly known as the continuum fitting method. \texttt{nkbb} models the thermal spectrum of an accretion disk in a non-Kerr spacetime geometry as measured by the distant observer~\cite{Zhou:2019fcg}. The peak of the thermal spectrum of the accretion disk with the inner edge at ISCO radius lies in the soft X-ray band ($ \sim 0.1-10$~keV) in the case of a stellar-mass black hole ($M \sim 10 M_{\odot}$) and in the UV band ($\sim 1-100$~eV) in the case of a supermassive black hole ($M \sim 10^5 - 10^9 M_{\odot}$)\footnote{Due to this, the continuum fitting method is normally more suitable for stellar-mass black holes rather than for supermassive black holes, as dust absorption in the UV band complicates the measurement in the case of supermassive black holes.}. A spectral hardening factor $f_{\rm col}$ is introduced in order to account for non-thermal effects, which are primarily due to electron scattering in the disk atmosphere~\cite{Davis:2004jf,Davis:2006bk,Davis:2018hlj}. The color temperature is defined as $T_{\rm col} (r) = f_{\rm col} T_{\rm eff}$. We can write the specific intensity of the thermal radiation, emitted in the rest-frame of the gas, in the units of $\rm {erg~s^{-1}~cm^{-2}~str^{-1}~Hz^{-1}}$ as~\cite{Zhou:2019fcg,Li:2004aq}
\begin{equation}
\label{LTS}
    I_{e} (\nu_{e}) = \frac{2 h \nu_{e}^3}{c^2} \frac{1}{f_{\rm col}^{4}} \frac{\Upsilon}{{\rm exp}(\frac{h \nu_{e}}{k_{\rm B} T_{\rm col}}) - 1},
\end{equation}
where $\nu_{e}$ is the emitted frequency of the photon, $h$ is the Planck's constant, $k_{\rm B}$ represents the Boltzmann constant, and $\Upsilon$ is a factor that depends upon the photon's emission angle, say $\xi$, with respect to the normal of the disk. $\Upsilon = 1$ (isotropic emission) and $\Upsilon = 1/2 + 3/4~{\rm cos}~\xi$ (limb-darkening emission) are the two popular options~\cite{Li:2004aq}.

Having determined the local thermal spectrum, the next step is to project the flux onto the observer's sky. This is accomplished by using the Cunningham transfer function approach~\cite{Cunningham:1975zz} (for more details, see Appendix~\ref{numerical_technique} and references therein). The transfer function serves as an integration kernel and takes into account all relativistic effects of the spacetime (gravitational redshifts, Doppler boosting, light bending)~\cite{Bambi:2016sac,Bambi:2017khi,Dauser:2010ne}. The transfer functions are computed using the ray-tracing technique in which photons are fired perpendicular to the observer's sky and are tracked until they hit the accretion disk. The trajectory of each photon in the curved spacetime is determined by solving the geodesic equations with the fourth-order Runge-Kutta (RK-4) method with adaptive step size. Our general relativistic ray-tracing code is described in Refs.~\cite{Abdikamalov:2019yrr,Ayzenberg:2018jip,Gott:2018ocn}. We compute the transfer functions for 200 radii of the disk, ranging from the disk inner edge to $10^6~M$. For each radius,  47 values of the transfer function are computed corresponding to different $g^*$ values. The $g^*$ values are distributed unevenly, with a narrow spacing close to 0 and 1 and widening towards 0.5. The advantage of the transfer function approach is two-fold: 1) it separates the calculation of relativistic effects from that of the local microphysics, and 2) it can be calculated over a range of parameters of spacetime and stored in a FITS (Flexible Image Transport System) table. The latter allows us to perform only the integration during the X-ray data fitting of a source, rather than running the full ray-tracing code, which is too slow.

\begin{figure}[tbp]
\centering 
\includegraphics[width=0.65\textwidth,trim=0 0 0 0,clip]{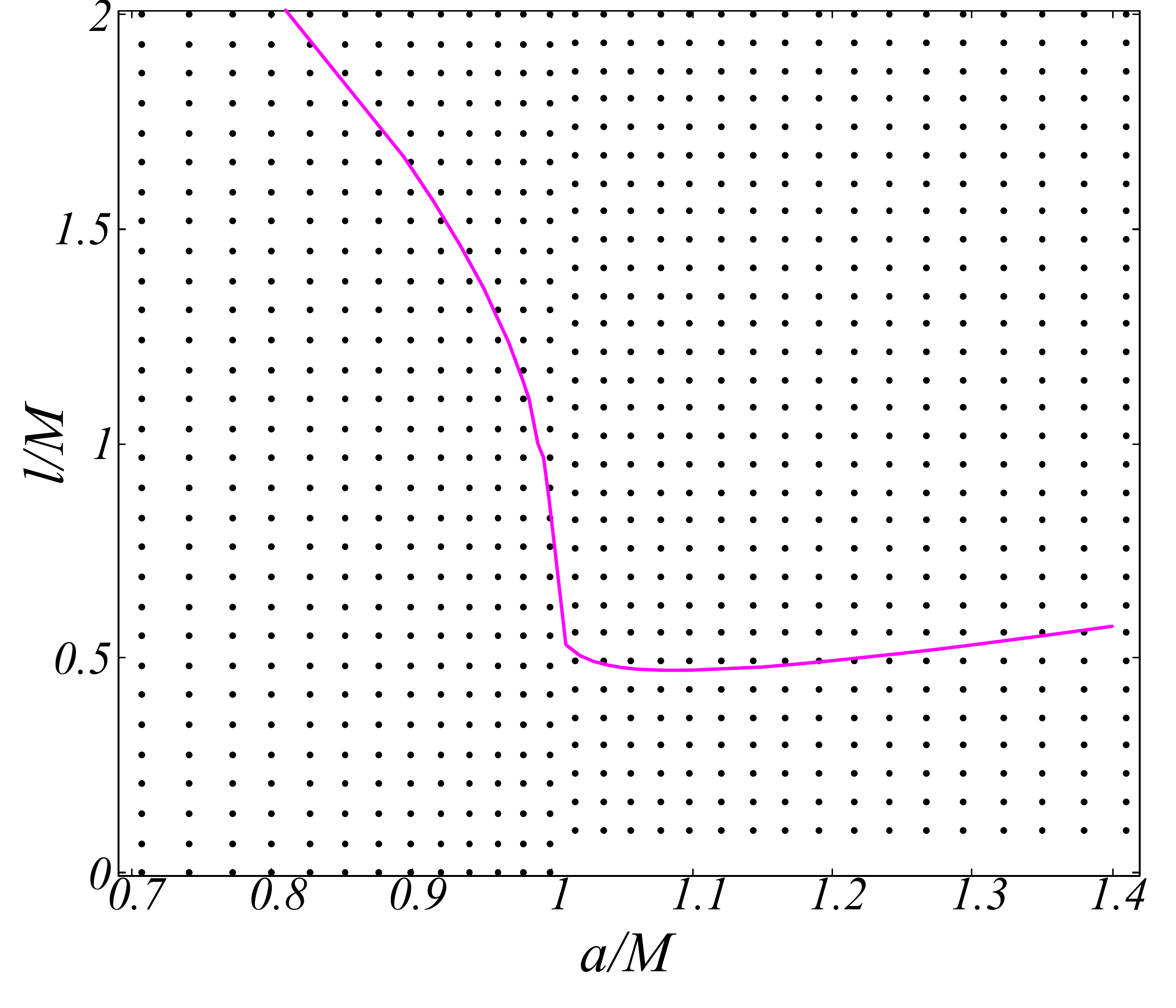}
\caption{Grid spanning the spin parameter $a/M$ and the regularization parameter $l/M$ of the spacetime, for which the transfer functions are computed and stored in a FITS file. The magenta curve represents the case of a wormhole with $l = r_{\rm ISCO}$ of the accretion disk. Note that, in order to avoid the spacetimes close to those with a naked singularity, we keep $l_{\rm min} > 0$, say $l_{\rm min} = 0.1~M$, for $a/M > 1$.}
\label{grid0}
\end{figure}

We compute the transfer functions for each grid point in the three-dimensional space spanned by ($a/M, l, i$). The ranges of $a/M$ and $l$ are, respectively, 
\begin{equation}
\label{spin_range}
     0.7 \le a/M \le 1.4
\end{equation}
\begin{equation}
\label{l_range}
     l = 
     \begin{cases}
        l_{\rm min} = 0 \le l \le l_{\rm max} = 2~M & \text{if } a/M \le 1 \\
        l_{\rm min} = 0.1~M \le l \le l_{\rm max} = 2~M & \text{if } a/M > 1
     \end{cases} .
\end{equation}
The grid points are shown in Figure~\ref{grid0}. The $l$ points are taken to be distributed evenly, whereas the $a/M$ points are slightly denser close to 1 because, in this region, the ISCO radius changes rapidly with the spin, requiring more sample points. We consider only $a/M \ge 0.7$ because our primary goal is to test the Kerr hypothesis, which require the analysis of the spectra of fast-rotating black holes~\cite{Dauser:2013xv}; thus, spin parameter less than 0.7 can be ignored as we will analyze the data of a fast-rotating black hole. This also saves computation time, but if we wanted to construct a full model we should cover the whole parameter space. Note that we exclude the $l < 0.1~M$ region when $a/M > 1$ to avoid the spacetimes close to those with a naked singularity. The inclination angles are distributed evenly in the range of $ 0< \text{cos }i < 1$ with 22 gird points. Eventually we have (30, 30, 22) points in the three-dimensional grid, resulting in a total of 19800 configurations stored in a FITS file of 3.7 GB in size.

\begin{figure}[tbp]
\centering 
\includegraphics[width=0.95\textwidth,trim=0 0 0 0,clip]{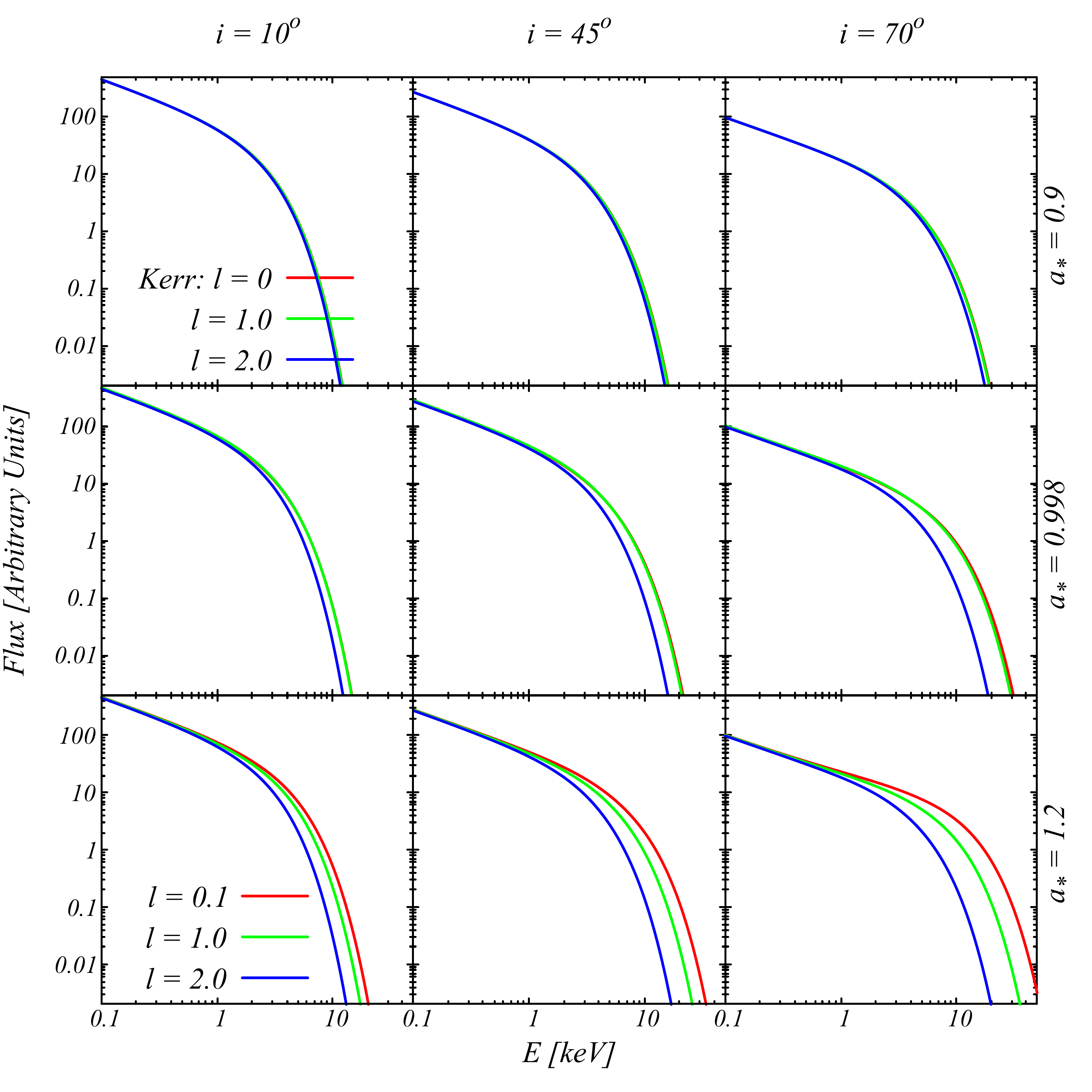}
\caption{Thermal spectra for spin $a_* = 0.9$ (top row), $a_* = 0.998$ (middle row), and $a_* = 1.2$ (bottom row). The inclination angle of the distant observer is taken as $i = 10^{\circ}$ (left column), $i = 45^{\circ}$ (middle column), and $i = 70^{\circ}$ (right column). The fixed parameters are as follows: torque at the inner edge of the accretion disk $\eta = 0$, mass of the compact object $M = 5.0~M_\odot$, mass accretion rate $\Dot{M} = 1\times10^{18}$~g/s, source's distance $D = 2.4$~kpc, and color factor $f_{\rm col} = 1.7$. For all the thermal spectra, the disk emission is assumed to obey the limb darkening law ($lflag = 1$). $l$ is in the units of gravitational radius  $r_{\rm g} = M$.}
\label{thermalspectra}
\end{figure}

The model is characterized by a total of 7 parameters ($M, D, i, \dot{M}, a/M, l/M, f_{\rm col}, \Upsilon $). The impact of different values of parameters $M$, $D$, and $\dot{M}$ on the synthetic thermal spectrum has already been explored in Refs.~\cite{Tripathi:2021rqs,Zhou:2019fcg}. Therefore, in Figure~\ref{thermalspectra}, we only show the impact of the parameters $a/M$, $i$ and $l/M$. Interestingly, for $a/M = 0.9$, the thermal spectrum is almost insensitive to the parameter $l/M$ because, at a lower spin, the disk's radiative efficiency remains almost constant with $l/M$ (as can be seen from Figure~\ref{eta_contours}). As the spin increases, the impact of increasing value of the parameter $l/M$ becomes clear, and now its effect is to soften the thermal spectrum. This is because the disk's radiative efficiency decreases with increasing $l/M$ at a higher spin. The overall effect of increasing the viewing angle $i$ is to enhance the Doppler boosting effects which, in turn, hardened the thermal spectrum.

%%%%%%%%%%%%%%%%%%%%%%%%%%%%%%%%%%%%%%%%%%%%%%%%%%%%%

\subsection{Relativistic reflection spectrum model: {\tt relxill\_nk}}

X-ray reflection spectroscopy refers to the analysis of the reflection spectrum of an accreting black hole~\cite{Bambi:2020jpe}. The disk-corona model serves as the theoretical foundation for this method. The term ``corona'' is used to indicate a hot ($T_{\rm e} \sim 100$ keV), usually optically thin, cloud of electrons in the vicinity of the black hole disk system~\cite{Wilkins:2012zm,Bambi:2020jpe}. Although the exact shape of the corona is unknown at the moment, the most common choices include a point-like emitting source -- presumably the base of some jet -- on the black hole's rotation axis (named the ``lamppost corona''), a static/rotating ring-like/disk-like source~\cite{Riaz:2020svt}, and/or some atmosphere above and below the disk; see, e.g., Refs.~\cite{Wilkins:2015nfa,Gonzalez:2017gzu} for more details and more coronal geometries.

Thermal photons from the accretion disk interact with coronal electrons via inverse Compton scattering, resulting in a spectrum that is usually approximated well by a power-law with a high energy cutoff $E_{\rm cut}$. The latter is related to the coronal temperature and $E_{\rm cut} \approx 2-3~T_{\rm e}$, depending on the exact coronal properties. Some Comptonized photons illuminate the accretion disk, where they are reprocessed and re-emitted as a reflection component~\cite{Garcia:2013oma}. A reflection spectrum is characterized by several emission features, the most prominent of which is usually the Fe K$\alpha$ emission line at 6.4 keV for neutral or weakly ionized iron and up to 6.97 keV for H-like ionized iron, as well as a Compton hump peaking at 20-30 keV~\cite{Bambi:2020jpe}. The emission features appear to be narrow in the gas's rest frame, but they become skewed and broadened in the observer's sky due to spacetime relativistic effects (gravitational redshift, Doppler boosting, light bending). As a result, observations of the reflection spectrum, particularly of the Fe K$\alpha$ complex, serve as a useful tool to determine the properties of the black hole accretion disk system~\cite{Reynolds:2007rx,Bambi:2017iyh,Reynolds:2002np,Fabian:1989ej,Dauser:2010ne}.

One of the most advanced models for fitting the reflection data is \texttt{relxill\textunderscore nk}~\cite{Bambi:2016sac,Abdikamalov:2019yrr,Abdikamalov:2020oci}, which is a non-Kerr extension of the reflection model \texttt{relxill}~\cite{Dauser:2013xv,Garcia:2013lxa}. \texttt{relxill\textunderscore nk} is composed of two parts: the non-relativistic disk reflection model \texttt{xillver}~\cite{Garcia:2013oma} and the relativistic line convolution model \texttt{relline}~\cite{Dauser:2014jka,Dauser:2013xv,Dauser:2010ne}. The line convolution model, \texttt{relline}, takes the local reflection spectrum, produced by \texttt{xillver}, from each point on the disk and projects it onto the observer's sky, taking into account all the relativistic effects of the spacetime. The model \texttt{relxill\textunderscore nk} for the spacetime metric~\ref{line_element} is constructed in the same manner as the model \texttt{nkbb}, as described in Section~\ref{model_nkbb} and in Appendix~\ref{numerical_technique} (see also references therein), with the parameters $a/M$, $l/M$, and $i$ having the same ranges and grid points distribution. The only difference is that the transfer functions are now computed for 100 disk radii, from the disk inner edge to $1000~M$. This is because the thermal emission decreases more slowly with radius than the reflection emission, so for the latter it is enough to include in the calculations a smaller portion of the accretion disk.

In its basic version, the \texttt{relxill\textunderscore nk} model includes the following physical parameters: emissivity index of the disk irradiation profile $q$, inner edge of the accretion disk $r_{\rm in}$, outer edge of the accretion disk $r_{\rm out}$, spin of the black hole $a/M$, regularization parameter of the sapcetime $l/M$, observer's viewing angle $i$, photon index of the incident power-law spectrum on the disk $\Gamma$, cutoff energy of the power-law spectrum $E_{\rm cut}$, ionization of the disk\footnote{The ionization parameter of the disk is defined as $\xi = 4\pi F_{\rm x}/n_{\rm e}$, where $F_{\rm x}$ is the disk incident flux from the corona and $n_{\rm e}$ is the disk electron density~\cite{Garcia:2013oma} which is usaully fixed at $n_{\rm e} = 1.2\times 10^{15}~\text{cm}^{-3}$. Additionally, the model is capable of considering either variable ionization (variant called {\tt relxillion\_nk})~\cite{Abdikamalov:2021rty} or variable electron density (variant called {\tt relxilldgrad\_nk})~\cite{Abdikamalov:2021ues} as a function of the disk radius when necessary.} log$\xi$, and iron abundance of the disk in solar units $A_{\rm Fe}$. Note that the black hole mass is no longer a parameter because, unlike the thermal spectrum, the reflection spectrum is not explicitly mass dependent. In theory, the disk's emissivity profile should be determined by the geometry of the corona. Due to the fact that it is unknown at the moment, the precise emissivity profile is uncertain. The most popular choice is an arbitrary geometry, which approximates the emissivity profile with a single, or a broken power-law which can be given as

\begin{equation}
\label{em_profile}
     I (r) \propto 
     \begin{cases}
        \frac{1}{r^{q_{\rm in}}} & \text{if } r < r_{\rm br} \\
        \frac{1}{r^{q_{\rm out}}} & \text{if } r \ge r_{\rm br}
     \end{cases}
\end{equation}
where $q_{\rm in}$ in and $q_{\rm out}$ out denote the inner and outer emissivity indices, respectively, and $r_{\rm br}$ denotes the breaking radius. The single power-law emissivity profile is recovered if $q_{\rm in} = q_{\rm out}$.    
\begin{figure}[tbp]
\centering 
\includegraphics[width=0.95\textwidth,trim=0 0 0 0,clip]{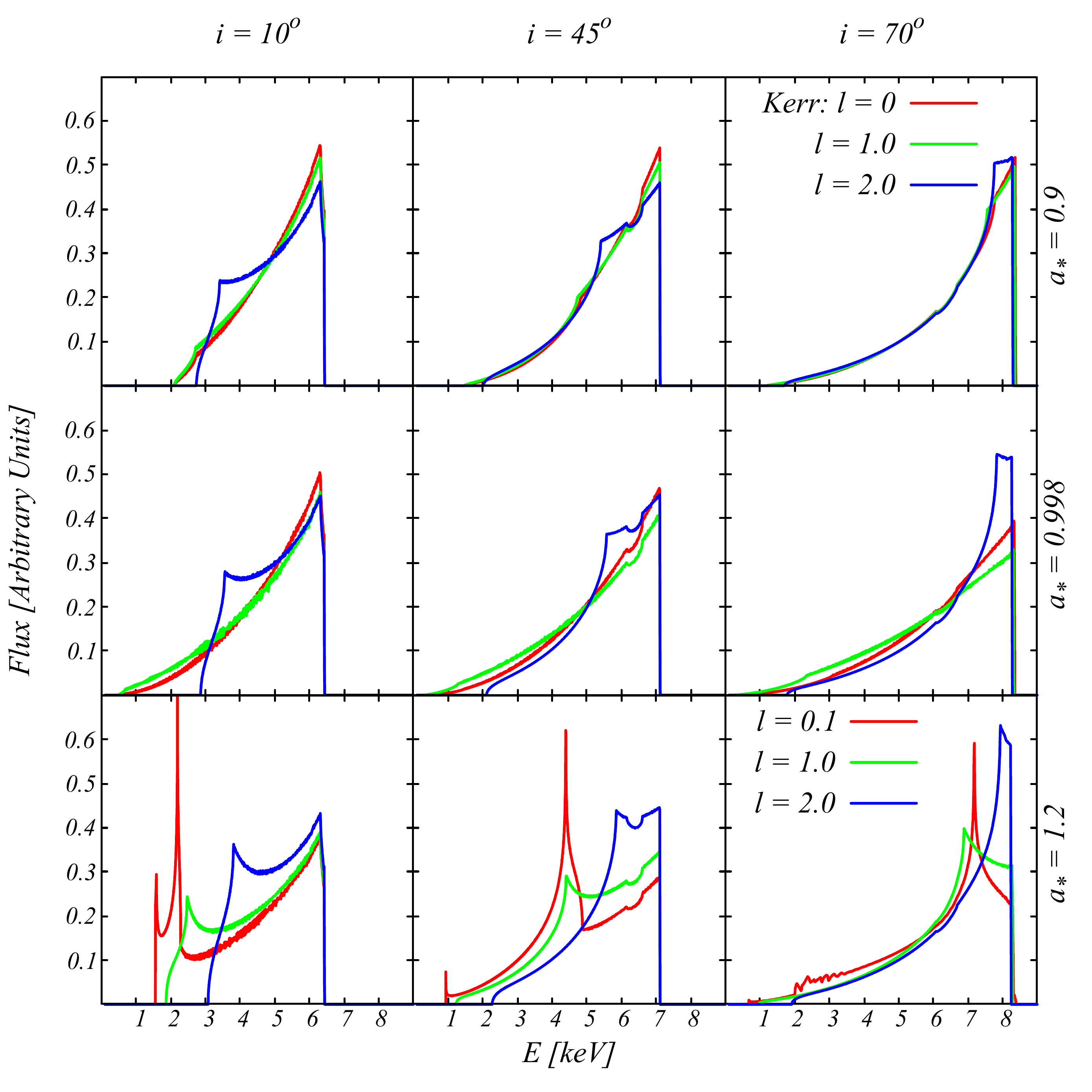}
\caption{Iron line profiles for different values of spin ($a/M$), observer's viewing angle ($i$), and regularization parameter ($l/M$) of the spacetime. The spin is varied along row direction as follows: top row ($a/M = 0.9$), middle row ($a/M = 0.998$), and bottom row ($a/M = 1.2$). The viewing angle is taken as $i = 10^{\circ}$ (left column), $i = 45^{\circ}$ (middle column), and $i = 70^{\circ}$ (right column). The emissivity profile of the accretion disk is set to be a single power-law with an index $q = 3$. $l$ is in the units of gravitational radius $r_{\rm g} = M$. }
\label{ironlines}
\end{figure}

Iron line profiles are shown in Figure~\ref{ironlines} for various values of spin ($a/M$), viewing angle ($i$) and spacetime regularization parameter ($l/M$). The disk's emissivity profile is set to be a single power-law with a $q = 3$ emissivity index. Each panel also contains the iron line for the Kerr case ($l = 0$), except for the case where $a/M > 1$, which avoids a naked singularity condition. The iron lines of $a/M \le 1$ will be discussed separately from those of $a/M > 1$. For $a/M \le 1$, the general trend is that the low energy tail of the iron line increases as $l = 0$ approaches $l = 1~M$. This is because, for a fixed spin, the ISCO radius decreases as $l$ increases (see Figure~\ref{ISCOcontours}), resulting in an increase in the number of highly redshifted photons in the low energy tail. This effect is amplified at a spin of $a/M = 0.998$, which further increases the low energy tail, because the ISCO radius decreases as spin increases from $a/M = 0.9$ to $a/M = 0.998$. The situation is different for $l = 2~M$ because it corresponds to a wormhole scenario with the ISCO radius inside the throat, truncating the disk's inner edge at $r_{\rm in} = l + 0.01~M$. As a result, in this case the iron lines lack the highly redshifted photons in the low energy tail. However, increasing the viewing angle increases the effect of Doppler boosting, which extends the low energy tail of the iron lines. We have wormhole scenarios in the case of $a/M = 1.2$. When $l = 0.1~M$, the disk's inner edge is at $r_{\rm ISCO}$, which results in an extreme redshift of the horn-like feature, which broadens and appears as an extended low energy tail at higher viewing angles due to Doppler boosting effects. For $l = 1~M$ and $l = 2~M$, the inner edge of the disk is truncated at $r_{\rm in} = l + 0.01~M$, resulting in a deficiency of highly redshifted photons in the low energy tail. However, we see some tail at higher viewing angles due to the dominating effect of Doppler boosting.

%%%%%%%%%%%%%%%%%%%%%%%%%%%%%%%%%%%%%%%%%%%%%%%%%%%%%

\section{Constraints from X-ray data}\label{sec:x}

In this section, we present a brief summary of the data reduction, data analysis, and constraints on spin ($a/M$) and the regularization parameter ($l/M$) of the spacetime acquired from the spectral fitting of the Galactic black hole candidate EXO 1846--031.

\subsection{Observation and data reduction}

EXO 1846--031 is a low-mass X-ray binary with a black hole. The source was discovered in 1985 by the \textsl{EXOSAT} mission~\cite{EXOSAT:mission}. After being quiescent for over two decades, it experienced a new outburst in 2019~\cite{Negoro:2019}. On 3 August 2019, the X-ray mission \textsl{NuSTAR}~\cite{NuSTAR:2013yza} observed this source (observation ID 90501334002) for a total exposure time of 22.2 ks~\cite{Draghis:2020ukh}; we analyze these data below. During this observation, the source was very bright. This observation was first analyzed in Ref.~\cite{Draghis:2020ukh}. Previous studies of this source identified a black hole with maximal spin and a high inclination angle of the accretion disk. Since the spectrum of the \textsl{NuSTAR} observation of EXO 1846--031 is simple and characterized by a prominent and broad iron line, this observation is ideally suited for testing the Kerr hypothesis~\cite{Tripathi:2020yts}.

We followed the data reduction procedure outlined in Ref.~\cite{Draghis:2020ukh}. We use the HEASOFT module {\tt nupipeline} to generate a cleaned event file from the raw data using NuSTARDAS and the CALDB 20220301 calibration database. The source spectra are derived from 180 arcseconds radius circular areas at the source's center on the Focal Plane Module (FPM) A and B detectors. The background regions are chosen to have the same size and to be at a sufficient distance from the source region along the diagonal so that the source photons' influence can be avoided. We generate source and background spectra, the response matrix file, and the ancillary file using the HEASOFT module {\tt nuproducts}. The spectra are grouped using {\tt grppha} so that each energy bin has at least 30 counts. Since the new CALDB fixes the issue in the 3-7 keV range, the fitting model does not require the {\tt nuMLIv1.mod} table~\cite{Draghis:2020ukh}.

%%%%%%%%%%%%%%%%%%%%%%%%%%%%%%%%%%%%%%%%%%%%%%%%%%%%%

\subsection{Spectral analysis}

For the spectral analysis, we use XSPEC 12.10.1s~\cite{xsepc:1996} adopting WILMS abundances~\cite{Wilms:2000ez} and VERNER cross-sections. We begin by fitting the data with an absorbed power-law model with a cutoff energy ({\tt tbabs$\times$cutoffpl} in XSPEC language). {\tt tbabs} models the Galactic absorption and has the hydrogen column density, $N_{\rm H}$, as a free parameter~\cite{Wilms:2000ez}. {\tt cutoffpl} models the power-law component of the corona and has as free parameters the photon index ($\Gamma$), the high energy cutoff ($E_{\rm cut}$), and its normalization. A floating constant is used to compensate for the minor difference between FPMA and FPMB detectors. The residuals below 4 keV indicate a disk thermal component, those between 6-7 keV indicate the presence of a broad iron line, and those between 10–30 keV indicate the presence of a Compton hump; the latter two are recognized as the reflection features. Therefore, we use {\tt diskbb}~\cite{Makishima:1986ap} and {\tt relxillion\_nk}~\cite{Abdikamalov:2021rty} to model these residuals. In XSPEC, the complete model can be expressed as
\begin{align*}
   { \tt const \times tbabs \times (cutoffpl + diskbb + relxillion\_nk). }
\end{align*}
{\tt diskbb} models the thermal component of the accretion disk in the Newtonian framework~\cite{Makishima:1986ap}. It has temperature at the inner edge of the disk ($T_{\rm in}$) as a free parameter in the fit. Since the thermal component is weak and fits well with a Newtonian model, a more complex model is no longer required. We model the reflection features in the spectrum with {\tt relxillion\_nk}: it is the {\tt relxill\_nk} variant that permits the radial variation of the ionization, as given by $\xi_{\rm in}$ (ionization at the inner edge of the disk) and the index $\alpha_{\xi}$ (which characterizes the radial profile). We first fit the data assuming a broken power-law for the emissivity profile. However, we get a very steep inner emissivity profile ($q_{\rm in} \approx 10$) and an almost flat outer emissivity profile ($q_{\rm out} \approx 0$). An almost flat emissivity profile may indicate an extended and quite uniform corona. However, this is not a self-consistent model, because the luminosity of the outer part of the disk goes like $r^2/r^{q_{\rm out}}$, so we must have $q_{\rm out} > 2$ in order to have a disk of finite luminosity. We thus refit the data assuming a twice broken power-law with four free parameters, namely the inner emissivity index $q_{\rm in}$, the breaking radius between the inner and the central part of the disk $r_{\rm br1}$, the central emissivity index $q_{\rm mid}$, and the breaking radius between the central and the outer part of the disk $r_{\rm br2}$. We freeze the outer emissivity index $q_{\rm out}$ to 3, which is the value expected at large radii for most coronal geometries. We fix the inner edge of the disk $r_{\rm in}$ to the ISCO and set the $r_{\rm out}$ parameter to 900~$M$.

\begin{figure}[tbh]
\centering 
\includegraphics[width=0.8\textwidth,trim=0 0 0 0,clip]{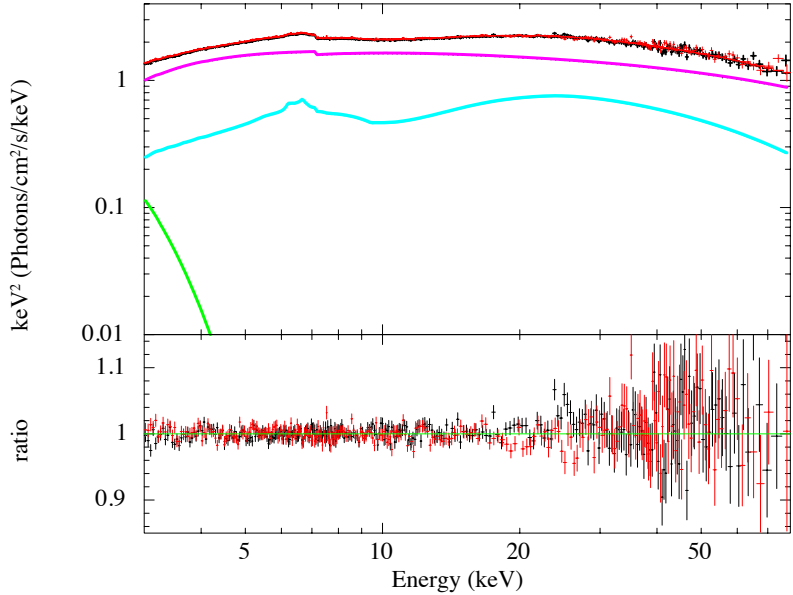}
\caption{Best-fit model (upper quadrant) and data to best-fit model ratio (lower quadrant) from the analysis of the 2019 \textsl{NuSTAR} observation of the X-ray binary EXO~1846--031. In the upper quadrant, the colors magenta, cyan, and green represent the power-law, relativistic reflection, and thermal components, respectively. Red crosses are used to indicate FPMA data and black crosses are for FPMB data.}
\label{best-fit-model-nk}
\end{figure}

\begin{figure}[tbp]
\centering 
\includegraphics[width=0.9\textwidth,trim=0 0 0 0,clip]{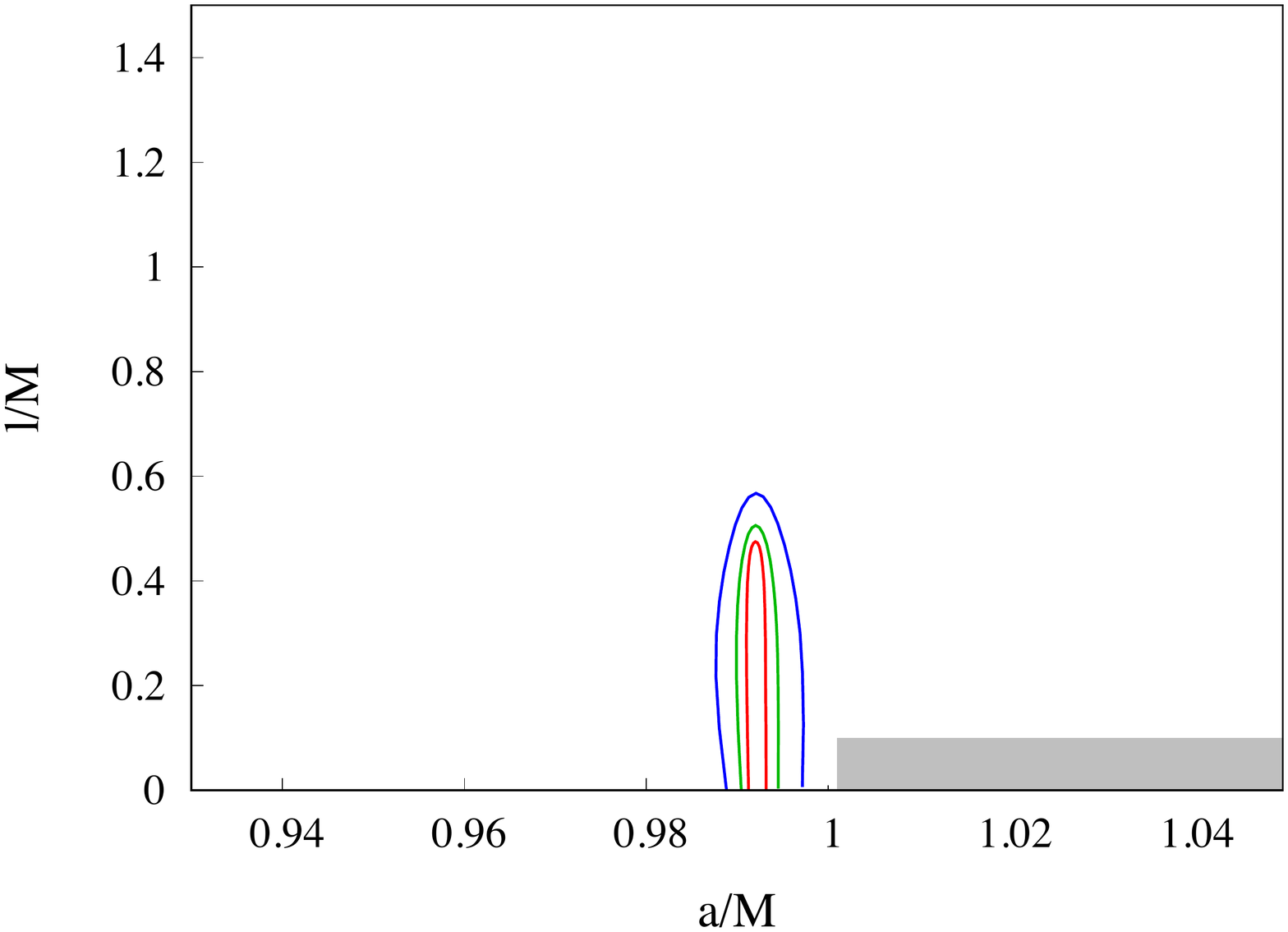}
\caption{Constraints on the spin ($a_*$) and regularization parameter ($l$) from the fit of the 2019 \textsl{NuSTAR} observation of the X-ray binary EXO~1846--031. The red, green, and blue curves represent, respectively, the 68\%, 90\%, and 99\% confidence level contours for two relevant parameters.}
\label{const}
\end{figure}

Table~\ref{t-fit} shows the parameter estimate of the best-fit. Figure~\ref{best-fit-model-nk} shows the best-fit model with the individual components and the residuals. Using the {\tt steppar} command in XSPEC, we determine the constraints on the spin and regularization parameters, as shown in Figure~\ref{const}. The red, green, and blue curves represent the 68\%, 90\%, and 99\% confidence level contours, respectively, for two relevant parameters.

\begin{table*}[tbh]
\centering
\caption{Summary of the analysis of the 2019 \textsl{NuSTAR} observation of the X-ray binary EXO~1846--031. The reported uncertainties correspond to the 90\% confidence level for one relevant parameter ($\Delta\chi^2 = 2.71$). $^\star$ indicates that the parameter is frozen in the fit. If there is no upper/lower uncertainty, it means that the 90\% confidence level reaches the upper/lower boundary of the parameter. This is the case for $q_{\rm in}$ and $q_{\rm mid}$, which are allowed to vary from 0 to 10, and for $l$, which must be positive.}
\label{t-fit}
{\renewcommand{\arraystretch}{1.3}
\begin{tabular}{lccccc}
\hline\hline
{\tt tbabs} &&&&& \\
$N_{\rm H}$ [$10^{22}$ cm$^{-2}$] & $$ &&& $$ & $6.50^{+0.09}_{-0.08}$ \\
\hline
{\tt diskbb}&&&&& \\
$kT_{\rm in}$ [keV] & $$ &&& $$ & $0.313^{+0.005}_{-0.010}$ \\
\hline
{\tt cutoffpl} \\
$\Gamma$ & $$ &&& $$ & $2.032^{+0.003}_{-0.017}$ \\
$E_{\rm cut}$ [keV] & $$ &&& $$ & $111^{+21}_{-2}$ \\
norm & $$ &&& $$ & $2.00^{+0.22}_{-0.07}$ \\
\hline
{\tt relxillion\_nk} &&&&& \\
$q_{\rm in}$ & $$ &&& $$ & $9.7_{-1.4}$ \\
$r_{\rm br1}$ [$r_{\rm g}$] & $$ &&& $$ & $6.03^{+0.50}_{-0.06}$ \\
$q_{\rm mid}$ & $$ &&& $$ & $0.13^{+0.08}$ \\
$r_{\rm br2}$ [$r_{\rm g}$]  & $$ &&& $$ & $538^{+80}_{-46}$ \\
$q_{\rm out}$ & $$ &&& $$ & $3^*$ \\
$i$ [deg] & $$ &&& $$ & $78.8^{+7.5}_{-1.2}$ \\
$a_*$ & $$ &&& $$ & $0.9920^{+0.0014}_{-0.0068}$ \\
$l$ [$r_{\rm g}$] & $$ &&& $$ & $0.44^{+0.05}$ \\
$\log\xi_{\rm in}$ [erg~cm~s$^{-1}$] & $$ &&& $$ & $3.03^{+0.08}_{-0.05}$ \\
$\alpha_{\xi}$ & $$ &&& $$ & $0.19^{+0.04}_{-0.03}$\\
$A_{\rm Fe}$ & $$ &&& $$ & $1.33^{+0.58}_{-0.12}$ \\
norm [$10^{-3}$] & $$ & && $$ & $0.60^{+0.60}_{-0.19}$ \\
\hline
$\chi^2/\nu$ &&&&& $\quad 2659.34/2598 \quad$ \\
&&&&& =1.02361 \\
\hline\hline
\end{tabular}}
\end{table*}

%%%%%%%%%%%%%%%%%%%%%%%%%%%%%%%%%%%%%%%%%%%%%%%%%%%%%

\section{Constraints from gravitational wave data}\label{sec:gw}

In this section, we continue our analysis using gravitational waves observed by the LIGO-Virgo Collaboration. The method we use for obtaining gravitational wave constraints on the metric in Eq.~\ref{line_element} is discussed in detail in \cite{Shashank:2021giy}. For obtaining the constraints using the gravitational wave data, we only use the inspiral phase of the merger which is modeled by the post-Newtonian (PN) framework \cite{Blanchet:2013haa}.

We consider the equatorial geodesics of massive particles around a black hole; i.e., $\dot\theta = 0$ and $\theta = \pi/2$. The independence of the metric on $t$ and $\phi$ implies the existence of two conserved quantities $E = -u_t$ and $L = u_\phi$, where $u^\mu$ is the four velocity of test particle around the black hole and $E$ and $L$ are the specific energy and angular momentum of the test particle. From the normalization condition of the four velocity $u^\mu u_\mu = -1$, one can write
\begin{equation}
    g_{rr}\dot r^2 = -1-g_{tt}\dot t^2-g_{\phi\phi}\dot \phi^2= V_\mathrm{eff}.
\end{equation}
The effective potential can be expanded in terms of $l^2$ away from the Kerr case (with $l^2\ll1$) and given by
\begin{equation}
    V_\mathrm{eff} = V_\mathrm{eff}^\mathrm{GR} + V_\mathrm{eff}^\mathrm{SV}+\mathcal{O}(l^4)
\end{equation}
Similarly, the energy and angular momentum for circular orbits can be rewritten as the term in Einstein gravity with a small contribution that depends on $l^2$
\begin{eqnarray}
	\begin{aligned}\label{eq:E_L_expansion}
		E &=& E^{\mathrm{GR}} + \delta E + \mathcal{O}(l^4) \, , \\
    		L &=& L^{\mathrm{GR}} + \delta L + \mathcal{O}(l^4) \, ,
	\end{aligned}
\end{eqnarray}
A modified Kepler's law in terms of the angular velocity $\Omega = d\phi/dt$, in the far-field limit reads
\be \label{eq:mod_kep_delta1}
    \Omega^2 = \frac{M}{r^3} \left[ 1 + \frac{3 M}{r} + \frac{9 M^2}{r^2} + \frac{M^2}{2r^2}l^2 + \mathcal{O} \left(l^4, \frac{M^3}{r^3} \right) \right] \, . \nonumber\\
\ee
where the regularization parameter $l^2$ enters at 2PN order \cite{Khan:2015jqa}. Under the leading order PN approximation for $l^2$, the spin effects can be ignored as they enter only at 3.5 PN or higher, effectively setting $a_* = 0$.

The total energy and the effective energy \cite{Damour:2000kk,Buonanno:1998gg} can be used to map our problem back to the two body problem. For circular orbits, the effective energy can be expressed in terms of the total energy ($E_T$) of the system, which is the energy of one body in the rest frame of the other body, \cite{Damour:2000kk,Buonanno:1998gg}:
\be
E_\mathrm{T} = m + E_\mathrm{b} = m [ 1 + 2 \eta (E_\mathrm{eff} - 1) ] \, ,    
\ee
where 
\be
E_\mathrm{eff} = g_{tt} \left( 1 + \frac{L^2}{r^2} \right)^{1/2} \, ,
\ee
where $E_\mathrm{b}$ represents the binding energy and $\eta = \mu/m$ is the symmetric mass ratio. Then the binding energy $E_\mathrm{b}$ is written as its term in Einstein gravity with a correction
\be
    E_\mathrm{b} = E_\mathrm{b}^\mathrm{GR} - \frac{\eta m^2}{r} \left[ l^2 \left( \frac{m}{r} \right)^2 + \mathcal{O} \left( l^4, \frac{m^3}{r^3} \right) \right] \, . \nonumber\\
\ee
The binding energy $E_\mathrm{b}$ is written in terms of the orbital frequency $\nu = \Omega/2\pi$ as the angular frequencies of the effective one-body problem and of the two-body problem are the same,
\be
    \frac{E_\mathrm{b}(\nu)}{\mu} = \frac{E_\mathrm{b}^\mathrm{GR}(\nu)}{\mu} - l^2(2 \pi m \nu)^{2}+\mathcal{O}\left[l^4,(2 \pi m \nu)^{8 / 3}\right] \, . \nonumber\\
\ee
The orbital phase is \cite{Shashank:2021giy,Cardenas-Avendano:2019zxd}
\be
    \phi(\nu) = \int^{\nu} \frac{1}{\dot{E}} \left(\frac{dE}{d \Omega}\right)  \Omega ~ d \Omega \, ,
\ee
where $\dot{E}$ represents the rate of change of binding energy caused by the emission of gravitational waves. X-ray constraints are only sensitive to the conservative sector and not on energy loss rate, and we are comparing gravitational wave results to those in the previous sections of this paper, so we consider only modifications on the conservative dynamics (binding energy) and we assume dissipative dynamics to be the same as in general relativity (a more detailed discussion is provided in Ref.~\cite{Cardenas-Avendano:2019zxd}). We only need to use the leading order 0PN for the change in binding energy \cite{Cardenas-Avendano:2019zxd,Blanchet:2013haa}; i.e., 
\be
\dot{E}_\mathrm{GR}^\mathrm{0PN} = -(32 / 5) \eta^{2} m^{2} r^{4} \Omega^{6} \, .
\ee
The orbital phase evolution is obtained as
\be
    \phi(\nu) = \phi_{\mathrm{GR}}(\nu)-\frac{25}{16 \eta}(2 \pi m \nu)^{-1 / 3} l^2+\mathcal{O}\left[l^4,(2 \pi m \nu)^{0}\right] \, , \nonumber\\
\ee
where
\be
\phi_{\mathrm{GR}}^\mathrm{0PN}(\nu)=-\frac{1}{32 \eta}(2 \pi m \nu)^{-5/3} \, .
\ee

The Fourier transform of $\phi$ in the stationary phase approximation \cite{Shashank:2021giy,Cardenas-Avendano:2019zxd} in leading order is
\be\label{eq:GW_phase_delta1}
    \Psi_{\mathrm{GW}}(f)=\Psi_{\mathrm{GW}}^{\mathrm{GR}}(f)-\frac{75}{32} u^{-1/3} \eta^{-4/5} l^2 + \mathcal{O}[l^4, u^0] \, , \nonumber\\
\ee
where
\be
\Psi_{\mathrm{GW}}^{\mathrm{GR}, 0 \mathrm{PN}}(f) = - \frac{3 u^{-5/3}}{128} \, 
\ee
and $u = \eta^{3/5} \pi m f$ and $f$ is the Fourier frequency.

Comparing Eq.~\ref{eq:GW_phase_delta1} to the parameterized post-Einsteinian (ppE) framework of Ref.~\cite{Yunes:2009ke}, $ \Psi_{\mathrm{GW}} = \Psi_{\mathrm{GW}}^{\mathrm{GR}} + \beta u^{b} $ and $b = -1/3$ at 2PN, we get
\be
\label{eq:beta_ppe}
    \beta = -\frac{75}{32} \eta^{-4/5} l^2 \, .
\ee
The ppE parameterization used by the LIGO-Virgo Collaboration is \cite{LalSuite}
\be \label{eq:beta_ligo}
    \beta = \frac{3}{128} \varphi_{4} \delta \varphi_{4} \eta^{-4/5} \, .
\ee
$\varphi_{4}$ is the PN phase at 2PN and has the form \cite{Khan:2015jqa}
\be
    \varphi_{4} = \frac{15293365}{508032}+\frac{27145 \eta}{504}+\frac{3085 \eta^{2}}{72} \, .
\ee
$\delta \varphi_4$ is the deviation from general relativity given as correction to the non-spinning portion of the PN phase \cite{LIGOScientific:2019fpa, LIGOScientific:2020tif}
\be
    \varphi_i \rightarrow (1 + \delta \varphi_i) \varphi_i \, .
\ee
Comparing Eqs. \ref{eq:beta_ppe} and \ref{eq:beta_ligo}, we find
\be\label{eq:d1-p4}
   l^2 = -\frac{1}{100} \varphi_{4} \delta \varphi_{4} \, .
\ee

The gravitational waveform depends on the conservative dynamics (here described by the binding energy) and on the dissipative dynamics (the energy loss rate). In our analysis we only consider modifications to the binding energy and we ignore any possible modification to the energy loss rate. We could not do otherwise, because in our model we have only the metric of the spacetime, while we do not have the full theory with the corresponding field equations, so we cannot calculate the energy loss rate, which is thus assumed to be the same as in general relativity. However, such a choice can also be justified as follows. As shown in Refs.~\cite{Yunes:2009ke,LIGOScientific:2019fpa}, gravitational wave data can constrain better lower or negative PN order effects. If modifications to the energy loss rate enter at a higher PN order than those to the binding energy, they can be neglected, because even if we include them we would obtain approximately the same constraints. If modifications to the energy loss rate enter at the same PN order as those to the binding energy, the constraints may change by a factor of order unity between the analysis with and without modifications to the energy flux. There are currently no known theories where there is a perfect or almost-perfect cancellation between the corrections to the binding energy and the energy loss rate~\cite{Cardenas-Avendano:2019zxd}. If modifications to the energy loss rate enter at a lower PN order than those to the binding energy, the actual constraints from the LIGO-Virgo data would be stronger than those obtained here without including modifications to the energy flux.

\begin{figure}[tbp]
\centering 
\includegraphics[width=\textwidth]{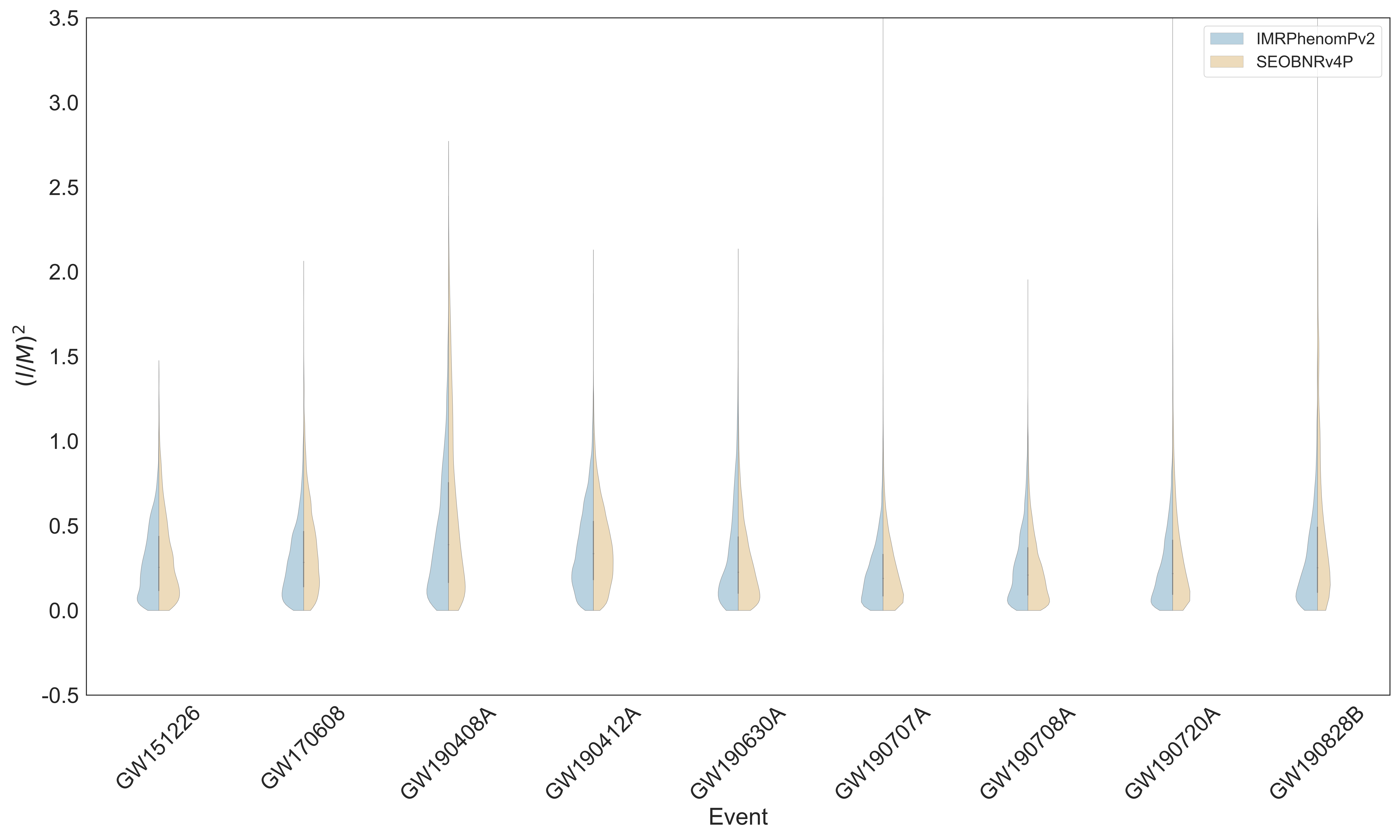}
\caption{Constraints on the metric parameter $(l/M)^2$ from the most significant events of GWTC-1 and GWTC-2. }
\label{grid}
\end{figure}

To obtain constraints on $l$, we fit the publicly available posterior samples released by the LIGO-Virgo Collaboration~\cite{datacat} and use the publicly available Markov Chain Monte Carlo (MCMC) sample, for the best fit model, inferred by two independent analysis ``Tests of General Relativity with Binary Black Holes from the second LIGO-Virgo Gravitational-Wave Transient Catalog - Full Posterior Sample Data Release"\footnote{\url{https://zenodo.org/record/5172704}}. The naming convention used here for the gravitational wave events is the same as used by the LIGO-Virgo Collaboration~\cite{datacat}. Table \ref{tab:delta1} and Fig.~\ref{grid} show the constraints on the mass normalized regularization parameter $l$ obtained using two waveform models, the phenomenological model IMRPhenomPv2 \cite{Hannam:2013oca} and the effective one body model SEOBNRv4P \cite{Bohe:2016gbl}. For the calculation of $l$, we use the individual red-shifted mass of the binaries and the deformation parameter $\delta\phi_4$. The data reported in Table~\ref{tab:delta1} shows the best fit value along with 90$\%$ confidence interval, with the lower limit set to zero\footnote{To obtain the constraints on $l$ from $l^2$ the data has been truncated at zero and hence the lower bound is set to zero}. Since our analysis is based on the PN expansion and an $(l/M)^2$ expansion, Table~\ref{tab:delta1} shows only the events with a total redshifted mass below 40~$M_\odot$ and for which the value of the median meets the condition $(l/M)^2 \le 0.5$. The 40~$M_\odot$ limit comes from the requirement that the LIGO-Virgo signal is dominated by the inspiral phase~\cite{Perkins:2022fhr}. In the next section, we will only consider the event providing the most stringent constraint, for which $(l/M)^2 \approx 0.2$ with the value of the median and $(l/M)^2 \approx 0.5$ with the value of the 90\% confidence level limit.

If a modified theory depends on a single parameter, in general the variation of that parameter will cause all PN terms to change. However, there is a hierarchy in the PN expansion, which is the crucial point to make our method both robust and reliable. For this reason we can constrain the parameter of the theory from the leading order modified PN term\footnote{We note that there are other methods to test general relativity with gravitational wave data that can measure multiple phase deformations simultaneously. One of these methods is the principal component analysis (PCA)~\cite{Saleem:2021nsb,Pai:2012mv}, which has its advantages and disadvantages (see the discussion in the conclusion section of Ref.~\cite{Perkins:2022fhr}.}. Including the modifications in the higher order PN terms into our analysis would not appreciably change the final constraint. The validity of single-parameter post-Einsteinian tests has been recently investigated and confirmed in Ref.~\cite{Perkins:2022fhr}, where the authors show that including higher order terms one gets a somewhat stronger constraint, not a weaker one, and the difference is small. This justifies our choice to consider only the leading order term in the ppE expansion and use the LIGO-Virgo Collaboration posterior distributions directly to constrain $l$.

We note that there are two caveats in this approach~\cite{Perkins:2022fhr}. First, the signal of the event must be dominated by the inspiral phase, where the PN expansion works. In practice, this condition limits this approach to binaries with sufficiently small total mass. For the characteristics of the LIGO-Virgo observatories, the total redshifted mass of the binary must be less than 40~$M_\odot$~\cite{Perkins:2022fhr}. The second caveat is that deviations from the general relativity signal have to be ``persistent''; that is, we may have problems to constrain gravity models in which modifications in the signal turn on suddenly (e.g., due to the activation of some additional field). In our analysis, we have simply ignored the events with total redshifted mass exceeding 40~$M_\odot$ to meet the first requirement. The second requirement is automatically satisfied because here we have only the background metric and we assume that the gravitational wave emission is the same as in general relativity.

\begin{table}
    \centering
    \def\arraystretch{1.5}
    \setlength{\tabcolsep}{10pt}
    %\begin{ruledtabular}
    \begin{tabular}{ccc}
        Event & $l/M$ (IMRPhenomPv2) & $l/M$ (SEOBNRv4P) \\
        \hline\hline
		GW151226 & $0.50 ^{+ 0.26 }$ & $0.51 ^{+ 0.31 }$ \\
		GW170608 & $0.49 ^{+ 0.26 }$ & $0.57 ^{+ 0.29 }$ \\		
		GW190408A & $0.58 ^{+ 0.39 }$ & $0.66 ^{+ 0.53 }$ \\
		GW190412A & $0.57 ^{+ 0.27 }$ & $0.59 ^{+ 0.26 }$ \\
		GW190630A & $0.48 ^{+ 0.38 }$ & $0.47 ^{+ 0.31 }$ \\
		GW190707A & $0.44 ^{+ 0.24 }$ & $0.44 ^{+ 0.28 }$ \\
		GW190708A & $0.47 ^{+ 0.28 }$ & $0.45 ^{+ 0.28 }$ \\
		GW190720A & $0.45 ^{+ 0.29 }$ & $0.48 ^{+ 0.38 }$ \\
		GW190728A & $0.41 ^{+ 0.36 }$ & $-$ \\
		GW190828B & $0.46 ^{+ 0.33 }$ & $0.57 ^{+ 0.52 }$ \\
		\hline \hline
    \end{tabular}
    %\end{ruledtabular}
    \caption{Median and the $90$th percentile of the parameter $l/M$ from the significant binary black hole events in GWTC-1 and GWTC-2 with the IMRPhenomPv2 and SEOBNRv4P waveform models. 
    We only consider events with a total redshifted mass below 40~$M_\odot$, since our method is valid only when the analyzed signal is dominated by the inspiral phase, and for which $(l/M)^2 \leq 0.5$, since our analysis is only valid in the regime of $(l/M)^2 \ll 1$.
    -- denotes that the data for the particular case is not available. Note the lower limit for the data is set to zero}
    \label{tab:delta1}
\end{table}

%%%%%%%%%%%%%%%%%%%%%%%%%%%%%%%%%%%%%%%%%%%%%%%%%%%%%

\section{Concluding remarks}\label{sec:dc}

In this work, we have considered the regular black hole metric recently proposed by Mazza, Franzin \& Liberati in Ref.~\cite{Mazza:2021rgq} and we have used the available X-ray and gravitational wave black hole data to measure the regularization parameter $l$.

In the case of X-ray data, we have analyzed the 2019 \textsl{NuSTAR} observation of the Galactic black hole EXO~1846--031. This spectrum is quite suitable for our test thanks to the high-quality data of the \textsl{NuSTAR} observation and the very strong and very broadened iron line of the source at that time. Our final result is
\be\label{eq-f-c-x}
l/M < 0.49 \quad \text{(90\% CL, only statistical error)} .
\ee
Our constraint does not include systematic uncertainties, which, however, should be subdominant~\cite{Tripathi:2020dni}. Some systematic uncertainties are related to simplifications in our theoretical model; in particular, we model the emissivity profile with a broken power-law, which is certainly an approximation of the actual emissivity profile, and our model ignore the returning radiation, which may have some effect in the spectra of fast-rotating black holes in which the emissivity profile of the inner part of the accretion disk is steep~\cite{Riaz:2020zqb}, as it is the case of our source.

We note that our constraint in \ref{eq-f-c-x} is probably close to the best that we can do with the available X-ray data from the analysis of the reflection spectrum of an accreting black hole. Somewhat better constraints could be obtained by joint analyses of the reflection features and the thermal spectrum of the disk, as done in Refs.~\cite{Tripathi:2021rqs,Tripathi:2020dni,Zhang:2021ymo}. To improve significantly this constraint, we need to wait for the next generation of X-ray missions, like \textsl{eXTP}~\cite{eXTP:2016rzs} or \textsl{Athena}~\cite{Nandra:2013jka}, which promise to provide unprecedented high-quality data of accreting black holes.

Concerning the gravitational wave constraints, as we can see from Table~\ref{tab:delta1} there are a few sources providing the bound $l/M < 0.8$ with some minor differences among different sources and between the models IMRPhenomPv2 and SEOBNRv4P. The most stringent constraint is from the gravitational wave event GW190707A which, in the worse case with the SEOBNRv4P model, turns out to be
\be\label{eq-f-c-gw}
l/M < 0.72 \quad \text{(90\% CL, only statistical error)}
\ee
Even in this case, the constraint does not include systematic uncertainties but, again, these should be subdominant with respect to the statistical error~\cite{Cardenas-Avendano:2019zxd}. The method employed has some simplifications, and in particular is based on an expansion and we do not include the impact of the spins. Like for the \textsl{NuSTAR} constraint, this is likely the best we can do with the available gravitational wave data. Somewhat better constraints can be obtained in the next cycles of observations of LIGO and Virgo if there are upgrading in the detectors. Substantial improvements of these constraints require the next generation of gravitational wave observatories.

Last, we note that there are already a few studies published in the literature on the observational constraints of the regularization parameter $l$ of this regular black hole metric.

In Ref.~\cite{Jiang:2021ajk}, following the method proposed in Ref.~\cite{Bambi:2013fea}, the authors estimate the regularization parameter $l$ from the analysis of the quasi-periodic oscillations (QPOs) in \textsl{RXTE} spectra of the Galactic black hole GRO~J1655--40 assuming the interpretation of Ref.~\cite{Motta:2013wga} of the QPO frequencies. The final measurement reported in Ref.~\cite{Jiang:2021ajk} is $l/M < 0.36$ at 68\% CL. We note that such a constraint is model-dependent, in the sense that it crucially depends on the assumption of the mechanism responsible for the generation of QPOs, which is currently unknown. However, it also shows the potentialities of QPO-based tests, in the sense that once the origin of QPOs will be understood well we can perform quite precise tests of the Kerr metric.

In Ref.~\cite{Shaikh:2021yux}, the authors try to constraint the regularization parameter $l$ from the shadow image of the supermassive black hole M87$^*$ released by the Event Horizon Telescope Collaboration in April 2019. Requiring that the deviation from circularity of the black hole shadow is less than 10\% and assuming that the black hole spin is in the range $0.5 < a_* < 0.94$ (which is the range inferred by the Event Horizon Telescope Collaboration for a Kerr black hole), they constrain $l$ in term of the black hole horizon, so their measurement is not directly comparable to ours. However, the constraint is roughly an order of magnitude weaker than those in \ref{eq-f-c-x} and \ref{eq-f-c-gw}, which is consistent with the difference of the constraining power between X-ray reflection spectroscopy and black hole shadow found in previous studies for other deformation parameters (see, e.g., Ref.~\cite{Tripathi:2020yts}).

The impact of the regularization parameter $l$ on strong gravitational lensing is studied in Ref.~\cite{Islam:2021ful}. However, the authors conclude that it is not possible constrain $l$ with the available data.

%%%%%%%%%%%%%%%%%%%%%%%%%%%%%%%%%%%%%%%%%%%%%%%%%%%%%

\appendix
%%%% APPENDIX%%%
\section{Numerical technique}\label{numerical_technique}

Our models \texttt{nkbb} and \texttt{relxill\textunderscore nk} are constructed by employing the Cunningham transfer function approach~\cite{Cunningham:1975zz}. We follow the same strategy as described in Refs.~\cite{Bambi:2016sac,Bambi:2017khi,Abdikamalov:2019yrr,Dauser:2010ne}; however, we briefly review it here for the reader's convenience. We consider the observer at spatial infinity $r = +\infty$, making an inclination angle $i$ between the observer's line of sight and the black hole's spin angular momentum. We can write the observed flux, in the units of ${\rm erg~s^{-1}~cm^{-2}~Hz^{-1}}$, as~\cite{Abdikamalov:2019yrr,Bambi:2017khi}
\begin{equation}
\label{flux}
 F_{o}(\nu _{o}) = \frac{1}{D^2} \int I_{o}(\nu_{o}, X, Y) dXdY = \frac{1}{D^2} \int g^3 I_{e} (\nu_{e}, r_{e}, \theta_{e}) dXdY ,     
\end{equation}
where $I_e$ and $I_o$ refer to, respectively, the specific intensities of radiation emitted in the rest-frame of the gas and detected by the distant observer. $\nu_o$ and $\nu_{e}$ represent, respectively, the photon's frequency observed and emitted in the gas rest-frame. $X$ and $Y$ are, respectively, the Cartesian coordinates of the accretion disk in the observer's plane. $D$ is the distance to the observer. $I_{e}$ and $I_{o}$ are related through the Liouville's theorem~\cite{Lindquist:1966igj}, $I_{e}/\nu_{e}^3 = I_{o}/\nu_{o}^3$, and $g$ is the redshift factor which can be written as
\begin{equation}
    g = \frac{\nu_{o}}{\nu_{e}} = \frac{(p_{a} u^{a})_{o}}{(p_{b} u ^{b})_{e}},
\end{equation}
where $p_{a}$ is the photon's conjugate momentum, which turns out to be $p_{a} = (-E^{\gamma}, p_{r}, p_{\theta}, L_{z}^{\gamma})$ for a stationary and axisymmetric spacetime, and $u_{e}^a$ and $u_{o}^a$ are, respectively, the 4-velocities of the emitter and the observer. For a stationary observer, we can write $u_{o}^{a} = (1,0,0,0)$ and hence, $(p_{a} u^a)_{o} = -E^{\gamma}$. The 4-velocity of the emitter turn out to be
\begin{equation}
    u_{e}^{a} =u^{t}_{e} (1, 0,0, \Omega),   
\end{equation}
where $\Omega$ is the angular velocity of the gas in the accretion disk (given in Eq.~\ref{angular_velocity}). By following the normalization condition, $g_{\mu \nu} u_{e}^{\mu} u_{e}^{\nu} = -1$, we can express $u^t_{e} = \dot{t}$ as
\begin{equation}
    \dot{t} = \frac{1}{\sqrt{-(g_{tt} + 2 g_{t \phi}\Omega + g_{\phi \phi}\Omega^2)}}. 
\end{equation}
Now, we can write $(p_{a} u^{a})_{e} = \dot{t} (-E^{\gamma} + \Omega L_{z}^{\gamma})$ and finally the redshift factor as
\begin{equation}
\label{redshfit_factor}
    g = \frac{\sqrt{-(g_{tt} + 2 g_{t \phi}\Omega + g_{\phi \phi}\Omega^2)}}{1 - \Omega b}, 
\end{equation}
where $b = L_{z}^{\gamma} / E^{\gamma}$. 

In eq.~\ref{flux}, $\theta_e$ is the photon's emission angle with respect to the disk's normal -- which is important since the emission from the disk's surface is non-isotropic~\cite{Svoboda:2009tp,Svoboda:2012cy,Garcia:2013lxa,Tripathi:2020cje} -- given by
\begin{equation}
    {\rm cos}~\theta_{e} = g \sqrt{g^{\theta \theta}} \frac{p_{\theta}^e}{p_{t}^e},
\end{equation}
where $g$ is again the redshift factor (Eq.~\ref{redshfit_factor}) and $p_{a}^e$ is the conjugate momentum of the photon at the emission in the accretion disk.

In terms of the transfer function, one can perform the integration over the disk in lieu of the observer's sky. The observed flux can be expressed as
\begin{equation}
    \label{trf_flux}
    F_{o}(\nu _{o}) = \frac{1}{D^2} \int_{r_{\rm in}} ^{r_{\rm out}} \int _{0} ^{1} \pi r_{e} \frac{g^2}{\sqrt{ g^* (1 - g^*)}} f(g^*, r_{e}, i) I_{e}(\nu_{e}, r_{e}, \theta_{e}) dg^* dr_{e},  
\end{equation}
where $r_{e}$ is the radial coordinate of the emission point in the disk. $r_{\rm in}$ and $r_{\rm out}$ are, respectively, the inner and the outer edges of the disk. $f(g^*, r_{e}, i)$ is the transfer function given by
\begin{equation}
\label{trf}
    f(g^*, r_{e}, i) = \frac{1}{\pi r_{e}} g \sqrt{g^*(1 - g^*)} \left| \frac{\partial(X,Y)}{\partial (g^*, r_{e})} \right|, 
\end{equation}
where $\left| \partial(X,Y) / \partial (g^*, r_{e}) \right|$ is the Jacobian of the coordinate transformation from $(X,Y)$ to $(r_{e}, g^*)$. $g^*$, in Eqs.~\ref{trf_flux} and~\ref{trf}, is the relative redshift factor, varies from $g^* = 0$ to $g^* = 1$ for a given $r_{e}$, defined as 
\begin{equation}
 g^* = \frac{g - g_{\rm min}}{g_{\rm max} - g_{\rm min}},     
\end{equation}
where $g_{\rm max} (r_{e}, i)$ and $g_{\rm min} (r_{e}, i)$ are, respectively, the maximum and the minimum redshift factor for photons at a given emission $r_{e}$ on the disk and detected by the distant observer with viewing angle $i$. 

For a given $r_{e}$ and $i$, in general, the transfer function, turn out to be a closed curve parameterized by $g^*$, having two branches, $f^{(1)} (g^*, r_{e}, i)$ and $f^{(2)} (g^*, r_{e}, i)$, connected at the points $g^* = 0$ and $g^* = 1$. The observed flux in Eq.~\ref{trf_flux} can be re-expressed as~\cite{Bambi:2017khi,Abdikamalov:2019yrr}
\begin{align}
\label{trf_2branches}
     F_{o} (\nu_{o}) = \frac{1}{D^2} \int_{r_{\rm in}} ^{r_{\rm out}} \int _{0} ^{1} \pi r_{e} \frac{g^2}{\sqrt{ g^* (1 - g^*)}} f^{(1)}(g^*, r_{e}, i) I_{e}(\nu_{e}, r_{e}, \theta_{e}^{(1)}) dg^* dr_{e} \\+
     \frac{1}{D^2} \int_{r_{\rm in}} ^{r_{\rm out}} \int _{0} ^{1} \pi r_{e} \frac{g^2}{\sqrt{ g^* (1 - g^*)}} f^{(2)}(g^*, r_{e}, i) I_{e}(\nu_{e}, r_{e}, \theta_{e}^{(2)}) dg^* dr_{e}   
\end{align}
where $\theta^{(1)}_{e}$ and $\theta^{(2)}_{e}$ are, respectively, the emission angle of photon along the transfer branch 1 and 2.

%%%%%%%%%%%%%%%%%%%%%%%%%%%%%%%%%%%%%%%%%%%%%%%%%%%%%

\acknowledgments

We thank Alejandro~C\'ardenas-Avenda\~no for useful discussions and suggestions.
This work was supported by the National Natural Science Foundation of China (NSFC), Grant No. 11973019, the Natural Science Foundation of Shanghai, Grant No. 22ZR1403400, the Shanghai Municipal Education Commission, Grant No. 2019-01-07-00-07-E00035, and Fudan University, Grant No. JIH1512604.
S.S. also acknowledges support from the China Scholarship Council (CSC), Grant No.~2020GXZ016646.
R.R. also acknowledges support from the Shanghai Government Scholarship (SGS).

%%%%%%%%%%%%%%%%%%%%%%%%%%%%%%%%%%%%%%%%%%%%%%%%%%%%%

\end{document}